\documentclass[a4paper,11pt]{article}
\usepackage{multirow}
\pdfoutput=1

\usepackage{bbold}
\usepackage{ulem}
\usepackage{graphicx}
\usepackage{tikz}
\usepackage{tikz-cd}
\usetikzlibrary{decorations.pathmorphing}

\usepackage{todonotes}

\usepackage{jheppub} 

\newcommand{\rmT}{\scalebox{0.5}{$\mathrm{T}$}}

\usepackage{ytableau}
\usepackage{esint}
\usepackage{xcolor}
\definecolor{green2}{rgb}{0.0, 0.7, 0.0}

\title{\boldmath
M2-brane matrix models and Fermi gas on affine ABCD quivers
}

\author[a,b]{Taro Kimura}
\emailAdd{taro.kimura@ube.fr}
\affiliation[a]{
Université Bourgogne Europe, CNRS, IMB UMR 5584, 21000 Dijon, France
}

\affiliation[b]{Institut Universitaire de France (IUF), France}

\author[c,d]{and Tomoki Nosaka}
\emailAdd{nosaka@simis.cn}

\affiliation[c]{
Center for Mathematics and Interdisciplinary Sciences, Fudan University, Shanghai, 200433, China
}

\affiliation[d]{
 Department of Physics and Center for Field Theory and Particle Physics, Fudan University, 2005 Songhu Road, Shanghai 200438, China
 }

\abstract{
We study a new class of M2-brane matrix models associated with affine $BC$ and twisted affine $AD$ quivers. 
These models exhibit the characteristic $N^{3/2}$ scaling of the free energy in the large $N$ limit.
We establish a Fermi gas formalism and demonstrate that their grand canonical sum is given by a Fredholm Pfaffian, which is in parallel with the affine $D$ quiver matrix models.
The resulting Fermi gas density matrix shows a rational dependence on the canonical position/momentum operators in addition to the standard hyperbolic dependence.
This leads to unconventional sub-leading corrections, which obstruct the Airy function structure in the $1/N$ expansion of the free energy observed in the previously studied M2-brane matrix models.
}

\allowdisplaybreaks
\begin{document}
\maketitle

\ytableausetup{boxsize=1.5mm}

\flushbottom

\section{Introduction and summary}

There is a class of three-dimensional ${\cal N}=2$ supersymmetric gauge theories that can be interpreted as worldvolume theories of M2-branes in M-theory.
Typical examples include the three-dimensional ${\cal N}=4$ $\mathrm{U}(N)$ super Yang--Mills theory with one adjoint and fundamental hypermultiplets called Atiyah--Drinfeld--Hitchin--Manin (ADHM) theory \cite{deBoer:1996mp,deBoer:1996ck}, and the $\text{U}(N)_k\times \text{U}(N)_{-k}$ circular quiver Chern--Simons theory called Aharony--Bergman--Jafferis--Maldacena (ABJM) theory \cite{Hosomichi:2008jd,Aharony:2008ug}.
These models admit brane realizations in type IIA/IIB string theory, which can be uplifted to the systems of M2-branes probing certain background geometries.
By extending these type IIA/IIB brane setups, one can construct a wide variety of M2-brane theories, including those with longer circular quivers, i.e., affine $A$ type quivers, non-uniform ranks of the gauge groups and/or the orthogonal/symplectic gauge groups instead of the unitary gauge groups.
Applying the supersymmetric localization technique \cite{Kapustin:2009kz}, the $S^3$ partition function of these theories reduces to the finite (${\cal O}(N)$) dimensional integrals, which we shall call ``M2-brane matrix models''.
In the large $N$ limit, the free energy of these M2-brane matrix models scales in $N$ universally as $N^{3/2}$ \cite{Herzog:2010hf}, which is consistent with the effective degrees of freedom on $N$ M2-branes predicted based on the gauge/gravity duality~\cite{Klebanov:1996ag}.

Conversely, the $N^{3/2}$ scaling of the free energy can be regarded as a model to be a theory of $N$ M2-branes.
Hence, from the perspective of seeking new classes of M2-brane theories,
it would be useful to consider various deformations of the three-dimensional supersymmetric gauge theory or the M2-brane matrix model itself, preserving the $N^{3/2}$ scaling, which would not necessarily be realized in type IIA/IIB brane setups. 
In the large $N$ saddle point approximation \cite{Herzog:2010hf}, the $N^{3/2}$ scaling of the free energy relies on the cancellation of the ${\cal O}(N^2)$ interaction terms over the entire quiver diagram, which is dubbed long-range force cancellation. 
In \cite{Gulotta:2011vp} it was found that the long-range force cancellation is achieved not only for the circular quiver matrix models but also for the affine $DE$ type quivers with the ranks of the gauge groups scaled by the (co)marks of the corresponding affine root system.
We can also realize the long-range force cancellation for more general quivers, such as affine $BCF$ and twisted affine $ADE$ by assigning the orthogonal/symplectic gauge groups to a part of the quiver nodes or adding adjoint hypermultiplets \cite{Gulotta:2011vp,Amariti:2019pky}.

For the ABJM theory, the partition function can be written in the Fermi gas formalism \cite{Marino:2011eh}, which reveals the all order $1/N$ perturbative corrections to the free energy and also provides effective tools to study the $1/N$ non-perturbative corrections \cite{Calvo:2012du,Putrov:2012zi,Boldis:2026zqu}.
Besides powerful tool, the connection between the M2-brane matrix model and a quantum curve through the Fermi gas formalism is essential for understanding of topological string free energies in the non-perturbative corrections \cite{Hatsuda:2013oxa,Grassi:2014zfa,Honda:2014npa} and the $q$-Painlev\'e equation governing the finite $N$ exact values of the matrix model \cite{Grassi:2014uua,Bonelli:2017gdk}.
However, the Fermi gas formalism has not yet been established for all the M2-brane matrix models described above.
While the formalism covers the circular quiver theories with O/USp gauge groups \cite{Mezei:2013gqa,Okuyama:2015auc,Honda:2015rbb,Moriyama:2015asx,Moriyama:2016xin,Okuyama:2016xke,Moriyama:2016kqi} and/or rank deformations \cite{Awata:2012jb,Matsumoto:2013nya}, it has not yet been extended to other types of quivers except affine $D$ type quivers \cite{Assel:2015hsa,Moriyama:2015jsa,Kubo:2024raz}.

In this paper, we propose a new type of M2-brane matrix models of affine $BC$ and twisted affine $AD$ quivers.
In contrast to the known construction involving O/USp gauge groups depending on the node~\cite{Gulotta:2012yd}, we instead consider a node-dependent $\beta$-deformation.
We in particular assign $\beta = 2$ for ``bulk'' nodes, and $\beta = 1$ and $\beta = 4$ for short and long ``boundary'' nodes.
See \cite{Babinet:2022nxg} for finite quiver studies.
This realization is motivated by the (fractional) quiver W-algebra formalism of five-dimensional $\mathcal{N}=1$ gauge theory that involves the node-dependent $\Omega$-background~\cite{Kimura:2015rgi,Kimura:2016dys,Kimura:2017hez}.
Our matrix models are indeed obtained from the five-dimensional quiver gauge theory through the reduction. 

In the following, the gauge group rank is taken to be proportional to the mark of the corresponding root system. 
In this case, we show that the grand partition functions are given by Fredholm Pfaffains, i.e., square root of Fredholm determinants, establishing the Fermi gas formalism for these affine quiver matrix models.
We then study the large $N$ asymptotics of the grand partition functions based on the established Fermi gas method. 
Under the Chern--Simons level balance condition, we obtain the $N^{3/2}$ scaling of the free energy. 
We also find peculiar sub-leading contributions to the free energy which are absent in the known matrix models.
These sub-leading corrections spoil the Airy function structure
of the $1/N$ perturbative free energy which holds for the affine $AD$ quiver matrix models written in the Fermi gas formalism \cite{Marino:2011eh,Assel:2015hsa,Moriyama:2015jsa} (see also \cite{Mezei:2013gqa}) and also proposed to hold for several deformations of these matrix models which are not written in the Fermi gas formalism \cite{Kubo:2024qhq,Geukens:2024zmt,Bobev:2025ltz}.

The rest of the paper is organized as follows.
In section \ref{sec:aff_quiver_matrix_model} we define the quiver matrix models associated with the (twisted) affine Dynkin quiver diagrams.
In section \ref{sec_Fremigas} we construct the Fermi gas formalism for each of the quiver matrix models, where the integration measure associated with each node is unspecified.
In section \ref{sec_N32byFermigas} we choose the Fresnel measure with the Chern--Simons levels subject to the level balance condition \eqref{level_balance}, and determine the coefficient of $N^{3/2}$ in the large $N$ limit of the free energy by using the Fermi gas formalism.
In section \ref{sec_derivationQMM} we discuss the derivation of the $\beta$-deformed quiver matrix models studied in this paper.
In section \ref{sec_discussion} we summarize our findings and discuss possible future directions.
We summarize our convention for the one-dimensional quantum mechanics in appendix \ref{app_1dQM} and list the determinant formulas in appendix \ref{app_formulas}, which are used in the derivation of the Fermi gas formalism.
In appendix \ref{app_sec_D(h)(2)r+1} we derive the Fermi gas formalism for $D^{(2)}_{r+1}$ type quiver matrix model with odd ranks.
The Fermi gas formalism of this case has a slightly different structure from the other setups which is analogous to the open string formalism of the rank deformed affine $AD$ type quiver matrix models, and hence we skip in the main text for the presentation of the unified treatment of all the other setups.

\section{Affine quiver matrix model}\label{sec:aff_quiver_matrix_model}

Let $\Gamma=(\Gamma_0,\Gamma_1)$ be an affine Dynkin quiver diagram with $\Gamma_0=\{0,1,\cdots,r\}$ the set of nodes, and $\Gamma_1$ the set of edges.
From the simple roots $(\alpha_s)_{s\in\Gamma_0}$ associated with the nodes, we define the symmetrization $(b_{st})_{s,t\in\Gamma_0}$ of the Cartan matrix
\begin{align}
b_{st}=(\alpha_s,\alpha_t),
\end{align}
which is related to the Cartan matrix $(c_{st})_{s,t\in\Gamma_0}$ and the diagonal matrix $(d_{st})_{s,t\in\Gamma_0}$
\begin{align}
c_{st}=\frac{2(\alpha_s,\alpha_t)}{(\alpha_t,\alpha_t)}=(\alpha_s,\alpha_t^\vee),\quad
d_{st}=\frac{(\alpha_s,\alpha_s)\delta_{st}}{2}
\end{align}
as
\begin{align}
b=cd.
\end{align}

We define the quiver matrix model associated with each affine Dynkin diagram by using this matrix $b$ as
\begin{multline}
Z(\Gamma)=\int \prod_{s\in\Gamma_0} \left[\frac{1}{N_s!}\prod_{i=1}^{N_s} dx^{(s)}_i
\mu^{(s)}(x^{(s)}_i)
\prod_{i<j}^{N_s}\biggl|2\sinh\frac{x^{(s)}_i-x^{(s)}_j}{2}\biggr|^{b_{ss}} \right] \\
\times \prod_{s<t}\prod_{i=1}^{N_s}\prod_{j=1}^{N_t}\Bigl(2\cosh\frac{x^{(s)}_i-x^{(t)}_j}{2}\Bigr)^{b_{st}}.
\label{GammatoZrule}
\end{multline}
Here the measure factors $\mu^{(s)}(x)$ and the ``rank'' variable $N_s$ of each node can be chosen arbitrarily.
However, if we denote $N_s=a_sN$ and require that the integrand of the matrix model \eqref{GammatoZrule} does not diverges as $e^{{\cal O}(N)x^{(s)}_i}$ as $x^{(s)}_i\rightarrow\infty$ in the large $N$ limit, which implies
\begin{align}
\sum_{t\in\Gamma_0}b_{st}a_t=0
\end{align}
for all $s\in\Gamma_0$, this fixes the ratios of the ranks $(a_s)_{s\in\Gamma_0}$ to be the Kac label of type $\Gamma$:
A null root in affine root systems is given by $\delta = \sum_{s \in \Gamma_0} a_s \alpha_s$. 
We denote by ``$0$'' the affine node, and we write $\tilde{\Gamma}_0 = \Gamma_0\setminus \{0\}$.
Then, we have $a_0 = 1$ and $a_s = (\delta,\omega_s^\vee)$ for $s \in \tilde{\Gamma}_0$, where $(\omega_s)_{s \in \tilde{\Gamma}_0}$ and $(\omega_s^\vee)_{s \in \tilde{\Gamma}_0}$ are the fundamental weights and coweights obeying $(\alpha_s,\omega_t^\vee) = (\omega_s,\alpha_t^\vee) = \delta_{st}$. For a finite part of the Cartan subalgebra $\tilde{\mathfrak{h}}$, we have $(\delta,\tilde{\mathfrak{h}}) = 0$.
We call a set of ranks parametrized by a single integer $N$ as $N_s=a_s N$ ``balanced'', and in this paper we focus only on the quiver matrix models \eqref{GammatoZrule} with balanced rank variables.

Later in section \ref{sec_N32byFermigas} we also restrict $\mu^{(s)}(x)$ to be the Fresnel measures $\mu^{(s)}(x)=e^{\frac{ik_s}{4\pi}x^2}$, with $k = (k_s)_{s \in \Gamma_0}$ satisfying the level balance condition
\begin{align}
\sum_{s\in\Gamma_0}a_sk_s=0.
\label{level_balance}
\end{align}
In general, this condition is described as follows.
Let $\sigma = (\sigma_s)_{s \in \tilde{\Gamma}_0} \in \mathbb{R}^r$, which is interpreted as an element of $\tilde{\mathfrak{h}}$.
Then, for simple roots $\alpha_s \in \mathfrak{h}^*$, we take $k_s = (\alpha_s,\sigma)$ for $s \in \Gamma_0$. 
Indeed, we have $a \cdot k = \sum_{s \in \Gamma_0} a_s (\alpha_s,\sigma) = (\delta,\sigma) = 0$ for $\sigma \in \tilde{\mathfrak{h}}$.

The level balance condition \eqref{level_balance} together with the balanced ranks ensures that the effective action
\begin{align}
S_\text{eff}(\Gamma)&=-\sum_{s\in\Gamma_0}\frac{ik_s}{4\pi}\sum_{i=1}^{a_sN}(x^{(s)}_i)^2-\sum_{s\in\Gamma_0}\sum_{i<j}^{a_sN}b_{ss}\log\biggl|2\sinh\frac{x^{(s)}_i-x^{(s)}_j}{2}\biggr|\nonumber \\
&\quad + \sum_{s<t}\sum_{i=1}^{a_sN}\sum_{j=1}^{a_tN}b_{st}\log\Bigl(2\cosh\frac{x^{(s)}_i-x^{(t)}_j}{2}\Bigr),
\label{Seff}
\end{align}
which is the logarithm of the integrand of each matrix model, scales in the large $N$ limit as $S_\text{eff}\sim N^{3/2}$ under the following configuration
\begin{align}
x^{(s)}_i=\sqrt{N} X(\sigma)+Y^{(s)}(\sigma),\quad \Bigl(\sigma=\frac{\sigma-1}{N_s-1}-\frac{1}{2}\Bigr)
\label{HKPTansatz}
\end{align}
with $X(\sigma)$ beging monotonically increasing piecewise smooth real ${\cal O}(N^0)$ function of $\sigma\in(-\frac{1}{2},\frac{1}{2})$ and $Y^{(s)}(\sigma)$ being smooth complex ${\cal O}(N^0)$ functions of $\sigma$.
This property of the effective action suggests that the ``free energy'' $-\log Z(\Gamma)$ of the matrix model \eqref{GammatoZrule} scales in the large $N$ limit as $-\log Z(\Gamma)\sim N^{3/2}$ \cite{Herzog:2010hf}, which we support via analysis through the Fermi gas formalism of the matrix models in section \ref{sec_N32byFermigas}.

First let us consider the $B^{(1)}_r$ quiver
\begin{align}
\begin{tikzpicture}[baseline=(current bounding  box.center)]
\draw(0,1.4) circle (0.5);
\node at (0,1.4) {$0$};
\draw(0,0) circle (0.5);
\node at (0,0) {$1$};
\draw (0.45,0.21)--(1.05,0.49);
\draw (0.45,1.19)--(1.05,0.91);
\draw(1.5,0.7) circle (0.5);
\node at (1.5,0.7) {$2$};
\draw (2,0.7)--(2.2,0.7);
\node at (2.6,0.69) {$\cdots$};
\draw (3,0.7)--(3.2,0.7);
\draw(3.7,0.7) circle (0.5);
\node at (3.7,0.7) {$r-1$};
\draw (4.19,0.8)--(4.91,0.8);
\draw (4.19,0.6)--(4.91,0.6);
\draw (4.43,0.9)--(4.67,0.7)--(4.43,0.5);
\draw(5.4,0.7) circle (0.5);
\node at (5.4,0.7) {$r$};
\end{tikzpicture},
\end{align}
where the norms of the simple roots $\alpha_s$, the Cartan matrix $(c_{st})_{s,t\in\Gamma_0}$, its symmetrization $(b_{st})_{s,t\in\Gamma_0}$ and the Kac labels $a_s$ are
\begin{subequations}
\begin{align}
&(\alpha_s,\alpha_s)=\begin{cases}
2&\quad (s=0,\cdots,r-1)\\
1&\quad (s=r)
\end{cases},\quad
(c_{st})_{s,t\in\Gamma_0}=
\begin{pmatrix}
2 &0 &-1& 0&      & &   &\\
0 &2 &-1& 0&      & &   &\\
-1&-1&2 &-1&      & &   &\\
 0& 0&-1& 2&-1    & &   &\\
  &  &  &  &\ddots& &   &\\
  &  &  &  &    -1& 2&-1&0\\
  &  &  &  &      &-1& 2&-2\\
  &  &  &  &      & 0&-1&2
\end{pmatrix},\\
&(b_{st})_{s,t\in\Gamma_0}=
\begin{pmatrix}
2 &0 &-1& 0&      & &   &\\
0 &2 &-1& 0&      & &   &\\
-1&-1&2 &-1&      & &   &\\
 0& 0&-1& 2&-1    & &   &\\
  &  &  &  &\ddots& &   &\\
  &  &  &  &    -1& 2&-1&0\\
  &  &  &  &      &-1& 2&-1\\
  &  &  &  &      & 0&-1&1
\end{pmatrix},\quad
a_s=\begin{cases}
1&\quad (s=0,1)\\
2&\quad (s=2,\cdots,r)
\end{cases}.
\end{align}
\end{subequations}
Therefore the quiver matrix model for the balanced ranks $N_s=a_sN$, $Z(B^{(1)}_r;N)$, is built according to the rule \eqref{GammatoZrule} as
\begin{align}
&Z(B^{(1)}_r;N)\nonumber \\
&=
\int \prod_{s=0}^{1}\frac{1}{N!}\prod_{i=1}^{N}\frac{dx^{(s)}_i}{2\pi}\mu^{(s)}(x^{(s)}_i)
\int \prod_{s=2}^{r}\frac{1}{(2N)!}\prod_{i=1}^{2N}\frac{dx^{(s)}_i}{2\pi}\mu^{(s)}(x^{(s)}_i)\nonumber \\
&\quad\times
\frac{
\prod_{s=0}^{1}\prod_{i<j}^{N}(2\sinh\frac{x^{(s)}_i-x^{(s)}_j}{2})^2
\prod_{s=1}^{r-1}\prod_{i<j}^{2N}(2\sinh\frac{x^{(s)}_i-x^{(s)}_j}{2})^2
\prod_{i<j}^{2N}|2\sinh\frac{x^{(r)}_i-x^{(r)}_j}{2}|
}{
\prod_{s=0}^1\prod_{i=1}^{N}\prod_{j=1}^{2N}2\cosh\frac{x^{(s)}_i-x^{(2)}_j}{2}\prod_{s=2}^{r-1}\prod_{i,j=1}^{2N}2\cosh\frac{x^{(s)}_i-x^{(s+1)}_j}{2}}.
\end{align}

For the $C^{(1)}_r$ quiver
\begin{align}
\begin{tikzpicture}[baseline=(current bounding  box.center)]
\draw(0,0) circle (0.5);
\node at (0,0.0) {$0$};
\draw (0.49,0.1)--(1.21,0.1);
\draw (0.49,-0.1)--(1.21,-0.1);
\draw (0.73,0.2)--(0.97,0)--(0.73,-0.2);
\draw(1.7,0) circle (0.5);
\node at (1.7,0) {$1$};
\draw (2.2,0)--(2.4,0);
\node at (2.8,-0.01) {$\cdots$};
\draw (3.2,0)--(3.4,0);
\draw(3.9,0) circle (0.5);
\node at (3.9,0) {$r-1$};
\draw (4.39,0.1)--(5.11,0.1);
\draw (4.39,-0.1)--(5.11,-0.1);
\draw (4.87,0.2)--(4.63,0)--(4.87,-0.2);
\draw(5.6,0) circle (0.5);
\node at (5.6,0) {$r$};
\end{tikzpicture},
\end{align}
the norms of the simple roots $\alpha_s$, the Cartan matrix $(c_{st})_{s,t\in\Gamma_0}$, its symmetrization $(b_{st})_{s,t\in\Gamma_0}$ and the Kac labels $a_s$ are
\begin{subequations}
\begin{align}
&(\alpha_s,\alpha_s)=\begin{cases}
2&\quad (s=1,\cdots,r-1)\\
4&\quad (s=0,r)
\end{cases},\quad
(c_{st})_{s,t\in\Gamma_0}=
\begin{pmatrix}
2 &-2& 0& 0&      & &   &\\
-1&2 &-1& 0&      & &   &\\
 0&-1&2 &-1&      & &   &\\
 0& 0&-1& 2&-1    & &   &\\
  &  &  &  &\ddots& &   &\\
  &  &  &  &    -1& 2&-1&0\\
  &  &  &  &      &-1& 2&-1\\
  &  &  &  &      & 0&-2&2
\end{pmatrix},\\
&(b_{st})_{s,t\in\Gamma_0}=
\begin{pmatrix}
4 &-2& 0& 0&      & &   &\\
-2&2 &-1& 0&      & &   &\\
 0&-1&2 &-1&      & &   &\\
 0& 0&-1& 2&    -1& &   &\\
  &  &  &  &\ddots& &   &\\
  &  &  &  &    -1& 2&-1&0\\
  &  &  &  &      &-1& 2&-2\\
  &  &  &  &      & 0&-2&4
\end{pmatrix},\quad
a_s=\begin{cases}
1&\quad (s=0,r)\\
2&\quad (s=1,\cdots,r-1)
\end{cases}.
\end{align}
\end{subequations}
Therefore the quiver matrix model for the balanced ranks $N_s=a_sN$, $Z(C^{(1)}_r;N)$, is built according to the rule \eqref{GammatoZrule} as
\begin{align}
&Z(C^{(1)}_r;N)\nonumber \\
&=
\int
\prod_{s=0,r}\frac{1}{N!}
\prod_{i=1}^{N}
\frac{dx^{(s)}_i}{2\pi}
\mu^{(s)}(x^{(s)}_i)
\prod_{s=1}^{r-1}\frac{1}{(2N)!}
\prod_{i=1}^{2N}
\frac{dx^{(s)}_i}{2\pi}
\mu^{(s)}(x^{(s)}_i)\nonumber \\
&\quad\times
\prod_{i<j}^{N}\Bigl(2\sinh\frac{x^{(0)}_i-x^{(0)}_j}{2}\Bigr)^4
\prod_{s=1}^{r-1}
\prod_{i<j}^{2N}\Bigl(2\sinh\frac{x^{(s)}_i-x^{(s)}_j}{2}\Bigr)^2
\prod_{i<j}^{N}\Bigl(2\sinh\frac{x^{(r)}_i-x^{(r)}_j}{2}\Bigr)^4\nonumber \\
&\quad\times
\frac{1
}{
\prod_{i=1}^{N}\prod_{j=1}^{2N}(2\cosh\frac{x^{(0)}_i-x^{(1)}_j}{2})^2
\prod_{s=1}^{r-2}
\prod_{i,j=1}^{2N}2\cosh\frac{x^{(s)}_i-x^{(s+1)}_j}{2}
}\nonumber \\
&\quad\times\frac{1}{
\prod_{i=1}^{2N}\prod_{j=1}^{N}(2\cosh\frac{x^{(r-1)}_i-x^{(r)}_j}{2})^2
}.
\end{align}

For the $A^{(2)}_{2r-1}$ quiver
\begin{align}
\begin{tikzpicture}[baseline=(current bounding  box.center)]
\draw(0,1.4) circle (0.5);
\node at (0,1.4) {$0$};
\draw(0,0) circle (0.5);
\node at (0,0) {$1$};
\draw (0.45,0.21)--(1.05,0.49);
\draw (0.45,1.19)--(1.05,0.91);
\draw(1.5,0.7) circle (0.5);
\node at (1.5,0.7) {$2$};
\draw (2,0.7)--(2.2,0.7);
\node at (2.6,0.69) {$\cdots$};
\draw (3,0.7)--(3.2,0.7);
\draw(3.7,0.7) circle (0.5);
\node at (3.7,0.7) {$r-1$};
\draw (4.19,0.8)--(4.91,0.8);
\draw (4.19,0.6)--(4.91,0.6);
\draw (4.67,0.9)--(4.43,0.7)--(4.67,0.5);
\draw(5.4,0.7) circle (0.5);
\node at (5.4,0.7) {$r$};
\end{tikzpicture},
\end{align}
the norms of the simple roots $\alpha_s$, the Cartan matrix $(c_{st})_{s,t\in\Gamma_0}$, its symmetrization $(b_{st})_{s,t\in\Gamma_0}$ and the Kac labels $a_s$ are
\begin{subequations}
\begin{align}
&(\alpha_s,\alpha_s)=\begin{cases}
2&\quad (s=0,\cdots,r-1)\\
4&\quad (s=r)
\end{cases},\quad
(c_{st})_{s,t\in\Gamma_0}=
\begin{pmatrix}
2 &0 &-1& 0&      & &   &\\
0 &2 &-1& 0&      & &   &\\
-1&-1&2 &-1&      & &   &\\
 0& 0&-1& 2&  -1  & &   &\\
  &  &  &  &\ddots& &   &\\
  &  &  &  &   -1 & 2&-1&0\\
  &  &  &  &      &-1& 2&-1\\
  &  &  &  &      & 0&-2&2
\end{pmatrix},\\
&(b_{st})_{s,t\in\Gamma_0}=
\begin{pmatrix}
2 &0 &-1& 0&      & &   &\\
0 &2 &-1& 0&      & &   &\\
-1&-1&2 &-1&      & &   &\\
 0& 0&-1& 2&-1    & &   &\\
  &  &  &  &\ddots& &   &\\
  &  &  &  & -1   & 2&-1&0\\
  &  &  &  &      &-1& 2&-2\\
  &  &  &  &      & 0&-2&4
\end{pmatrix},\quad
a_s=\begin{cases}
1&\quad (s=0,1,r)\\
2&\quad (s=2,\cdots,r-1)
\end{cases}.
\end{align}
\end{subequations}
Therefore the quiver matrix model for the balanced ranks $N_s=a_sN$, $Z(A^{(2)}_{2r-1};N)$, is built according to the rule \eqref{GammatoZrule} as
\begin{align}
&Z(A^{(2)}_{2r-1};N)\nonumber \\
&=
\int
\prod_{s=0,1,r}\frac{1}{N!}\prod_{i=1}^{N}\frac{dx^{(s)}_i}{2\pi}\mu^{(s)}(x^{(s)}_i)
\prod_{s=2}^{r-1}\frac{1}{(2N)!}\prod_{i=1}^{2N}\frac{dx^{(s)}_i}{2\pi}\mu^{(s)}(x^{(s)}_i)\nonumber \\
&\quad\times 
\prod_{s=0}^{1}\prod_{i<j}^{N}\Bigl(2\sinh\frac{x^{(s)}_i-x^{(s)}_j}{2}\Bigr)^2
\prod_{s=2}^{r-1}\prod_{i<j}^{2N}\Bigl(2\sinh\frac{x^{(s)}_i-x^{(s)}_j}{2}\Bigr)^2
\prod_{i<j}^{N}\Bigl(2\sinh\frac{x^{(r)}_i-x^{(r)}_j}{2}\Bigr)^4
\nonumber \\
&\quad\times\frac{1
}{
\prod_{s=0,1}
\prod_{i=1}^{N}\prod_{j=1}^{2N}2\cosh\frac{x^{(s)}_i-x^{(2)}_j}{2}
\prod_{s=2}^{r-2}\prod_{i,j=1}^{2N}2\cosh\frac{x^{(s)}_i-x^{(s+1)}_j}{2}
}\nonumber \\
&\quad\times\frac{1}{
\prod_{i=1}^{2N}\prod_{j=1}^{N}(2\cosh\frac{x^{(r-1)}_i-x^{(r)}_j}{2})^2}.
\end{align}

For the $A^{(2)}_{2r}$ quiver
\begin{align}
\begin{tikzpicture}[baseline=(current bounding  box.center)]
\draw(0,0) circle (0.5);
\node at (0,0.0) {$0$};
\draw (0.49,0.1)--(1.21,0.1);
\draw (0.49,-0.1)--(1.21,-0.1);
\draw (0.97,0.2)--(0.73,0)--(0.97,-0.2);
\draw(1.7,0) circle (0.5);
\node at (1.7,0) {$1$};
\draw (2.2,0)--(2.4,0);
\node at (2.8,-0.01) {$\cdots$};
\draw (3.2,0)--(3.4,0);
\draw(3.9,0) circle (0.5);
\node at (3.9,0) {$r-1$};
\draw (4.39,0.1)--(5.11,0.1);
\draw (4.39,-0.1)--(5.11,-0.1);
\draw (4.87,0.2)--(4.63,0)--(4.87,-0.2);
\draw(5.6,0) circle (0.5);
\node at (5.6,0) {$r$};
\end{tikzpicture},
\end{align}
the norms of the simple roots $\alpha_s$, the Cartan matrix $(c_{st})_{s,t\in\Gamma_0}$, its symmetrization $(b_{st})_{s,t\in\Gamma_0}$ and the Kac labels $a_s$ are
\begin{subequations}
\begin{align}
&(\alpha_s,\alpha_s)=\begin{cases}
2&\quad (s=1,\cdots,r-1)\\
1&\quad (s=0)\\
4&\quad (s=0,r)
\end{cases},\quad
(c_{st})_{s,t\in\Gamma_0}=
\begin{pmatrix}
2 &-1& 0& 0&      & &   &\\
-2&2 &-1& 0&      & &   &\\
 0&-1&2 &-1&      & &   &\\
 0& 0&-1& 2& -1   & &   &\\
  &  &  &  &\ddots& &   &\\
  &  &  &  &  -1  & 2&-1&0\\
  &  &  &  &      &-1& 2&-1\\
  &  &  &  &      & 0&-2&2
\end{pmatrix},\\
&(b_{st})_{s,t\in\Gamma_0}=
\begin{pmatrix}
1 &-1& 0& 0&      & &   &\\
-1&2 &-1& 0&      & &   &\\
 0&-1&2 &-1&      & &   &\\
 0& 0&-1& 2&   -1 & &   &\\
  &  &  &  &\ddots& &   &\\
  &  &  &  &    -1& 2&-1&0\\
  &  &  &  &      &-1& 2&-2\\
  &  &  &  &      & 0&-2&4
\end{pmatrix},\quad
a_s=\begin{cases}
1&\quad (s=r)\\
2&\quad (s=0,\cdots,r-1)
\end{cases}.
\end{align}
\end{subequations}
Therefore the quiver matrix model for the balanced ranks $N_s=a_sN$, $Z(A^{(2)}_{2r};N)$, is built according to the rule \eqref{GammatoZrule} as
\begin{align}
&Z(A^{(2)}_{2r};N)\nonumber \\
&=
\int
\frac{1}{N!}\prod_{i=1}^{N}\frac{dx^{(r)}_i}{2\pi}\mu^{(r)}(x^{(r)}_i)
\prod_{s=0}^{r-1}\frac{1}{2N!}\prod_{i=1}^{2N}\frac{dx^{(s)}_i}{2\pi}\mu^{(s)}(x^{(s)}_i)\nonumber \\
&\quad\times
\frac{
\prod_{i<j}^{2N}|2\sinh\frac{x^{(0)}_i-x^{(0)}_j}{2}|
\prod_{s=1}^{r-1}\prod_{i<j}^{2N}(2\sinh\frac{x^{(s)}_i-x^{(s)}_j}{2})^2
\prod_{i<j}^{N}(2\sinh\frac{x^{(r)}_i-x^{(r)}_j}{2})^4
}{
\prod_{s=0}^{r-2}\prod_{i,j=1}^{2N}2\cosh\frac{x^{(s)}_i-x^{(s+1)}_j}{2}
\prod_{i=1}^{2N}\prod_{j=1}^{N}(2\cosh\frac{x^{(r-1)}_i-x^{(r)}_j}{2})^2}.
\end{align}

For the $D^{(2)}_{r+1}$ quiver
\begin{align}
\begin{tikzpicture}[baseline=(current bounding  box.center)]
\draw(0,0) circle (0.5);
\node at (0,0.0) {$0$};
\draw (0.49,0.1)--(1.21,0.1);
\draw (0.49,-0.1)--(1.21,-0.1);
\draw (0.97,0.2)--(0.73,0)--(0.97,-0.2);
\draw(1.7,0) circle (0.5);
\node at (1.7,0) {$1$};
\draw (2.2,0)--(2.4,0);
\node at (2.8,-0.01) {$\cdots$};
\draw (3.2,0)--(3.4,0);
\draw(3.9,0) circle (0.5);
\node at (3.9,0) {$r-1$};
\draw (4.39,0.1)--(5.11,0.1);
\draw (4.39,-0.1)--(5.11,-0.1);
\draw (4.63,0.2)--(4.87,0)--(4.63,-0.2);
\draw(5.6,0) circle (0.5);
\node at (5.6,0) {$r$};
\end{tikzpicture},
\end{align}
the norms of the simple roots $\alpha_s$, the Cartan matrix $(c_{st})_{s,t\in\Gamma_0}$, its symmetrization $(b_{st})_{s,t\in\Gamma_0}$ and the Kac labels $a_s$ are
\begin{subequations}
\begin{align}
&(\alpha_s,\alpha_s)=\begin{cases}
2&\quad (s=1,\cdots,r-1)\\
1&\quad (s=0,r)
\end{cases},\quad
(c_{st})_{s,t\in\Gamma_0}=
\begin{pmatrix}
2 &-1& 0& 0&      & &   &\\
-2&2 &-1& 0&      & &   &\\
 0&-1&2 &-1&      & &   &\\
 0& 0&-1& 2&-1    & &   &\\
  &  &  &  &\ddots& &   &\\
  &  &  &  & -1   & 2&-1&0\\
  &  &  &  &      &-1& 2&-2\\
  &  &  &  &      & 0&-1&2
\end{pmatrix},\\
&(b_{st})_{s,t\in\Gamma_0}=
\begin{pmatrix}
1 &-1& 0& 0&      & &   &\\
-1&2 &-1& 0&      & &   &\\
 0&-1&2 &-1&      & &   &\\
 0& 0&-1& 2&  -1  & &   &\\
  &  &  &  &\ddots& &   &\\
  &  &  &  &   -1 & 2&-1&0\\
  &  &  &  &      &-1& 2&-1\\
  &  &  &  &      & 0&-1&1
\end{pmatrix},\quad
a_s=1,\quad (s=0,\cdots,r).
\end{align}
\end{subequations}
Therefore the quiver matrix model for the balanced ranks $N_s=a_sN$, $Z(D^{(2)}_{r+1};N)$, is built according to the rule \eqref{GammatoZrule} as
\begin{align}
&Z(D^{(2)}_{r+1};N)\nonumber \\
&=
\int \prod_{s=0}^r\frac{1}{N!}\prod_{i=1}^{N}\frac{dx^{(s)}_i}{2\pi}\mu^{(s)}(x^{(s)}_i)\nonumber \\
&\quad\times
\frac{
\prod_{i<j}^{N}|2\sinh\frac{x^{(0)}_i-x^{(0)}_j}{2}|
\prod_{s=1}^{r-1} \prod_{i<j}^{N}(2\sinh\frac{x^{(s)}_i-x^{(s)}_j}{2})^2
\prod_{i<j}^{N}|2\sinh\frac{x^{(r)}_i-x^{(r)}_j}{2}|
}{
\prod_{s=0}^{r-1}\prod_{i,j}^{N}2\cosh\frac{x^{(s)}_i-x^{(s+1)}_j}{2}
}.
\label{ZD}
\end{align}

\section{Fermi gas formalism}
\label{sec_Fremigas}

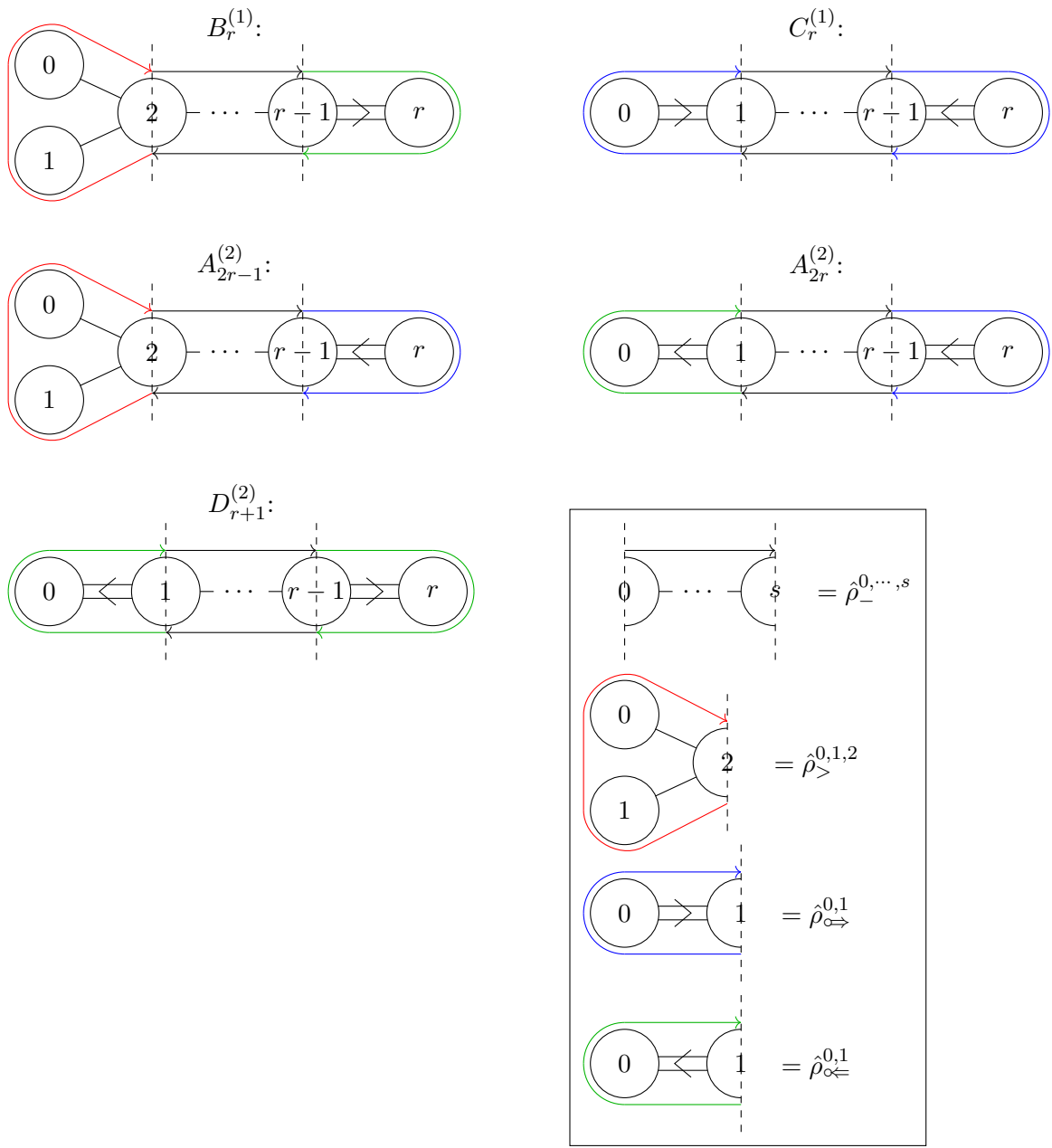
\begin{figure}
\begin{center}
\begin{tikzpicture}
%
%
\node at (2.7,2) {$B^{(1)}_r$:};
\draw(0,1.4) circle (0.5);
\node at (0,1.4) {$0$};
\draw(0,0) circle (0.5);
\node at (0,0) {$1$};
\draw (0.45,0.21)--(1.05,0.49);
\draw (0.45,1.19)--(1.05,0.91);
\draw(1.5,0.7) circle (0.5);
\node at (1.5,0.7) {$2$};
\draw (2,0.7)--(2.2,0.7);
\node at (2.6,0.69) {$\cdots$};
\draw (3,0.7)--(3.2,0.7);
\draw(3.7,0.7) circle (0.5);
\node at (3.7,0.7) {$r-1$};
\draw (4.19,0.8)--(4.91,0.8);
\draw (4.19,0.6)--(4.91,0.6);
\draw (4.43,0.9)--(4.67,0.7)--(4.43,0.5);
\draw(5.4,0.7) circle (0.5);
\node at (5.4,0.7) {$r$};
\draw [red] (0.254,-0.544) to [in=-90,out=205] (-0.6,0);
\draw [red] (0.254,-0.544)--(1.5,0.1);
\draw [red] (-0.6,0)--(-0.6,1.4);
\draw [red] (0.254,1.944) to [in=90,out=155] (-0.6,1.4);
\draw [red,->] (0.254,1.944)--(1.5,1.3);
\draw [dashed] (1.5,-0.3)--(1.5,1.7);
\draw [->] (1.5,1.3)--(3.7,1.3);
\draw [dashed] (3.7,-0.3)--(3.7,1.7);
\draw [green2] (3.7,1.3)--(5.4,1.3);
\draw [green2] (5.4,1.3) to [in=90,out=0] (6,0.7);
\draw [green2] (6,0.7) to [in=0,out=-90] (5.4,0.1);
\draw [green2,->] (5.4,0.1)--(3.7,0.1);
\draw [->] (3.7,0.1)--(1.5,0.1);
%
%
\node at (11.2,2) {$C^{(1)}_r$:};
\draw(8.4,0.7) circle (0.5);
\node at (8.4,0.7) {$0$};
\draw (8.89,0.8)--(9.61,0.8);
\draw (8.89,0.6)--(9.61,0.6);
\draw (9.13,0.9)--(9.37,0.7)--(9.13,0.5);
\draw(10.1,0.7) circle (0.5);
\node at (10.1,0.7) {$1$};
\draw (10.6,0.7)--(10.8,0.7);
\node at (11.2,0.69) {$\cdots$};
\draw (11.6,0.7)--(11.8,0.7);
\draw(12.3,0.7) circle (0.5);
\node at (12.3,0.7) {$r-1$};
\draw (12.79,0.8)--(13.51,0.8);
\draw (12.79,0.6)--(13.51,0.6);
\draw (13.27,0.9)--(13.03,0.7)--(13.27,0.5);
\draw(14,0.7) circle (0.5);
\node at (14,0.7) {$r$};
\draw [blue] (10.1,0.1)--(8.4,0.1);
\draw [blue] (8.4,0.1) to [in=-90,out=180] (7.8,0.7);
\draw [blue] (7.8,0.7) to [in=180,out=90] (8.4,1.3);
\draw [blue,->] (8.4,1.3)--(10.1,1.3);
\draw [dashed] (10.1,-0.3)--(10.1,1.7);
\draw [->] (10.1,1.3)--(12.3,1.3);
\draw [dashed] (12.3,-0.3)--(12.3,1.7);
\draw [blue] (12.3,1.3)--(14,1.3);
\draw [blue] (14,1.3) to [in=90,out=0] (14.6,0.7);
\draw [blue] (14.6,0.7) to [in=0,out=-90] (14,0.1);
\draw [blue,->] (14,0.1)--(12.3,0.1);
\draw [->] (12.3,0.1)--(10.1,0.1);
%
%
\node at (2.7,-1.5) {$A^{(2)}_{2r-1}$:};
\draw(0,-2.1) circle (0.5);
\node at (0,-2.1) {$0$};
\draw(0,-3.5) circle (0.5);
\node at (0,-3.5) {$1$};
\draw (0.45,-3.29)--(1.05,-3.01);
\draw (0.45,-2.31)--(1.05,-2.59);
\draw(1.5,-2.8) circle (0.5);
\node at (1.5,-2.8) {$2$};
\draw (2,-2.8)--(2.2,-2.8);
\node at (2.6,-2.81) {$\cdots$};
\draw (3,-2.8)--(3.2,-2.8);
\draw(3.7,-2.8) circle (0.5);
\node at (3.7,-2.8) {$r-1$};
\draw (4.19,-2.7)--(4.91,-2.7);
\draw (4.19,-2.9)--(4.91,-2.9);
\draw (4.67,-2.6)--(4.43,-2.8)--(4.67,-3);
\draw(5.4,-2.8) circle (0.5);
\node at (5.4,-2.8) {$r$};
\draw [red] (0.254,-4.044) to [in=-90,out=205] (-0.6,-3.5);
\draw [red] (0.254,-4.044)--(1.5,-3.4);
\draw [red] (-0.6,-3.5)--(-0.6,-2.1);
\draw [red] (0.254,-1.556) to [in=90,out=155] (-0.6,-2.1);
\draw [red,->] (0.254,-1.556)--(1.5,-2.2);
\draw [dashed] (1.5,-3.8)--(1.5,-1.8);
\draw [->] (1.5,-2.2)--(3.7,-2.2);
\draw [dashed] (3.7,-3.8)--(3.7,-1.8);
\draw [blue] (3.7,-2.2)--(5.4,-2.2);
\draw [blue] (5.4,-2.2) to [in=90,out=0] (6,-2.8);
\draw [blue] (6,-2.8) to [in=0,out=-90] (5.4,-3.4);
\draw [blue,->] (5.4,-3.4)--(3.7,-3.4);
\draw [->] (3.7,-3.4)--(1.5,-3.4);
%
%
%
\node at (11.2,-1.5) {$A^{(2)}_{2r}$:};
\draw(8.4,-2.8) circle (0.5);
\node at (8.4,-2.8) {$0$};
\draw (8.89,-2.7)--(9.61,-2.7);
\draw (8.89,-2.9)--(9.61,-2.9);
\draw (9.37,-2.6)--(9.13,-2.8)--(9.37,-3);
\draw(10.1,-2.8) circle (0.5);
\node at (10.1,-2.8) {$1$};
\draw (10.6,-2.8)--(10.8,-2.8);
\node at (11.2,-2.81) {$\cdots$};
\draw (11.6,-2.8)--(11.8,-2.8);
\draw(12.3,-2.8) circle (0.5);
\node at (12.3,-2.8) {$r-1$};
\draw (12.79,-2.7)--(13.51,-2.7);
\draw (12.79,-2.9)--(13.51,-2.9);
\draw (13.27,-2.6)--(13.03,-2.8)--(13.27,-3);
\draw(14,-2.8) circle (0.5);
\node at (14,-2.8) {$r$};
\draw [green2] (10.1,-3.4)--(8.4,-3.4);
\draw [green2] (8.4,-3.4) to [in=-90,out=180] (7.8,-2.8);
\draw [green2] (7.8,-2.8) to [in=180,out=90] (8.4,-2.2);
\draw [green2,->] (8.4,-2.2)--(10.1,-2.2);
\draw [dashed] (10.1,-3.8)--(10.1,-1.8);
\draw [->] (10.1,-2.2)--(12.3,-2.2);
\draw [dashed] (12.3,-3.8)--(12.3,-1.8);
\draw [blue] (12.3,-2.2)--(14,-2.2);
\draw [blue] (14,-2.2) to [in=90,out=0] (14.6,-2.8);
\draw [blue] (14.6,-2.8) to [in=0,out=-90] (14,-3.4);
\draw [blue,->] (14,-3.4)--(12.3,-3.4);
\draw [->] (12.3,-3.4)--(10.1,-3.4);
%
%
%
\node at (2.8,-5) {$D^{(2)}_{r+1}$:};
\draw(0,-6.3) circle (0.5);
\node at (0,-6.3) {$0$};
\draw (0.49,-6.2)--(1.21,-6.2);
\draw (0.49,-6.4)--(1.21,-6.4);
\draw (0.97,-6.1)--(0.73,-6.3)--(0.97,-6.5);
\draw(1.7,-6.3) circle (0.5);
\node at (1.7,-6.3) {$1$};
\draw (2.2,-6.3)--(2.4,-6.3);
\node at (2.8,-6.31) {$\cdots$};
\draw (3.2,-6.3)--(3.4,-6.3);
\draw(3.9,-6.3) circle (0.5);
\node at (3.9,-6.3) {$r-1$};
\draw (4.39,-6.2)--(5.11,-6.2);
\draw (4.39,-6.4)--(5.11,-6.4);
\draw (4.63,-6.1)--(4.87,-6.3)--(4.63,-6.5);
\draw(5.6,-6.3) circle (0.5);
\node at (5.6,-6.3) {$r$};
\draw [green2] (1.7,-6.9)--(0,-6.9);
\draw [green2] (0,-6.9) to [in=-90,out=180] (-0.6,-6.3);
\draw [green2] (-0.6,-6.3) to [in=180,out=90] (0,-5.7);
\draw [green2,->] (0,-5.7)--(1.7,-5.7);
\draw [dashed] (1.7,-7.3)--(1.7,-5.3);
\draw [->] (1.7,-5.7)--(3.9,-5.7);
\draw [dashed] (3.9,-7.3)--(3.9,-5.3);
\draw [green2] (3.9,-5.7)--(5.6,-5.7);
\draw [green2] (5.6,-5.7) to [in=90,out=0] (6.2,-6.3);
\draw [green2] (6.2,-6.3) to [in=0,out=-90] (5.6,-6.9);
\draw [green2,->] (5.6,-6.9)--(3.9,-6.9);
\draw [->] (3.9,-6.9)--(1.7,-6.9);
\draw (7.6,-5.1)--(12.8,-5.1)--(12.8,-14.4)--(7.6,-14.4)--cycle;
%
\draw ([shift={(8.4,-6.3)}]-90:0.5) arc[radius=0.5, start angle=-90, end angle= 90];
\node at (8.4,-6.3) {$0$};
\draw (8.9,-6.3)--(9.1,-6.3);
\node at (9.5,-6.31) {$\cdots$};
\draw (9.9,-6.3)--(10.1,-6.3);
%
\draw ([shift={(10.6,-6.3)}]90:0.5) arc[radius=0.5, start angle=90, end angle=270];
\node at (10.6,-6.3) {$s$};
\draw [->] (8.4,-5.7)--(10.6,-5.7);
\draw [dashed] (8.4,-7.3)--(8.4,-5.3);
\draw [dashed] (10.6,-7.3)--(10.6,-5.3);
\node at (11.9,-6.3) {$=\hat{\rho}^{0,\cdots,s}_-$};
\draw(8.4,-8.1) circle (0.5);
\node at (8.4,-8.1) {$0$};
\draw(8.4,-9.5) circle (0.5);
\node at (8.4,-9.5) {$1$};
\draw (8.85,-9.29)--(9.45,-9.01);
\draw (8.85,-8.31)--(9.45,-8.59);
\draw ([shift={(9.9,-8.8)}]90:0.5) arc[radius=0.5, start angle=90, end angle=270];
\node at (9.9,-8.8) {$2$};
\draw [red] (8.654,-10.044) to [in=-90,out=205] (7.8,-9.5);
\draw [red] (8.654,-10.044)--(9.9,-9.4);
\draw [red] (7.8,-9.5)--(7.8,-8.1);
\draw [red] (8.654,-7.556) to [in=90,out=155] (7.8,-8.1);
\draw [red,->] (8.654,-7.556)--(9.9,-8.2);
\draw [dashed] (9.9,-9.8)--(9.9,-7.8);
\node at (11.2,-8.8) {$=\hat{\rho}^{0,1,2}_>$};
\draw(8.4,-11) circle (0.5);
\node at (8.4,-11) {$0$};
\draw (8.89,-10.9)--(9.61,-10.9);
\draw (8.89,-11.1)--(9.61,-11.1);
\draw (9.13,-10.8)--(9.37,-11)--(9.13,-11.2);
\draw ([shift={(10.1,-11)}]90:0.5) arc[radius=0.5, start angle=90, end angle=270];
\node at (10.1,-11) {$1$};
\draw [blue] (10.1,-11.6)--(8.4,-11.6);
\draw [blue] (8.4,-11.6) to [in=-90,out=180] (7.8,-11);
\draw [blue] (7.8,-11) to [in=180,out=90] (8.4,-10.4);
\draw [blue,->] (8.4,-10.4)--(10.1,-10.4);
\draw [dashed] (10.1,-12)--(10.1,-10);
\node at (11.2,-11) {$=\hat{\rho}^{0,1}_{\circ\!\Rightarrow}$};
\draw(8.4,-13.2) circle (0.5);
\node at (8.4,-13.2) {$0$};
\draw (8.89,-13.1)--(9.61,-13.1);
\draw (8.89,-13.3)--(9.61,-13.3);
\draw (9.37,-13)--(9.13,-13.2)--(9.37,-13.4);
\draw ([shift={(10.1,-13.2)}]90:0.5) arc[radius=0.5, start angle=90, end angle=270];
\node at (10.1,-13.2) {$1$};
\draw [green2] (10.1,-13.8)--(8.4,-13.8);
\draw [green2] (8.4,-13.8) to [in=-90,out=180] (7.8,-13.2);
\draw [green2] (7.8,-13.2) to [in=180,out=90] (8.4,-12.6);
\draw [green2,->] (8.4,-12.6)--(10.1,-12.6);
\draw [dashed] (10.1,-14.2)--(10.1,-12.2);
\node at (11.2,-13.2) {$=\hat{\rho}^{0,1}_{\circ\!\Leftarrow}$};
\end{tikzpicture}
\caption{
The dictionary between affine quiver diagram and the building block operators \eqref{rhobb} for the Fermi gas density matrix $\hat{\rho}(\Gamma)$.
}
\label{rhovsquiver}
\end{center}
\end{figure}

In this section we show that the generating function of the quiver matrix models with respect to $N$
\begin{align}
\Xi(\Gamma;u)=\begin{cases}
\displaystyle \sum_{N=0}^\infty u^N Z(\Gamma;N)&\quad (\Gamma=B^{(1)}_r,C^{(1)}_r,A^{(2)}_{2r-1},A^{(2)}_{2r},\\
\displaystyle \sum_{N=0}^\infty u^N Z(D^{(2)}_{r+1};2N)&\quad (\Gamma=D^{(2)}_{r+1})
\end{cases}
\label{GC}
\end{align}
is written in a unified manner as the square root of a Fredholm determinant
\begin{align}
\Xi(\Gamma;u)=\sqrt{\operatorname{Det}(1+u\hat{\rho}({\Gamma}))}.
\label{Xi=sqrtDet}
\end{align}
Here $\operatorname{Det}(1+\hat{\cal O})$ is the Fredholm determinant $\operatorname{Det}(1+\hat{\cal O})=\exp\sum_{n=1}^\infty\frac{(-1)^{n-1}}{n}\operatorname{tr}\hat{\cal O}^n$, and $\hat{\rho}(\Gamma)$ is a one-dimensional quantum mechanical operator depending on the quiver as
\begin{subequations}
\begin{align}
&\hat{\rho}(B^{(1)}_r)=\hat{\rho}_{>}^{0,1,2}\hat{\rho}_-^{2,\cdots,r-1}\hat{\rho}^{r,r-1}_{\circ\!\Leftarrow}\hat{\rho}_-^{r-1,\cdots,2},\label{rhohatB} \\
&\hat{\rho}(C^{(1)}_r)=
\hat{\rho}_{\circ\!\Rightarrow}^{0,1}
\hat{\rho}_-^{1,\cdots,r-1}
\hat{\rho}_{\circ\!\Rightarrow}^{r,r-1}
\hat{\rho}_-^{r-1,\cdots,1},\label{rhohatC} \\
&\hat{\rho}(A^{(2)}_{2r-1})=
\hat{\rho}_{>}^{0,1,2}
\hat{\rho}_-^{2,\cdots,r-1}
\hat{\rho}_{\circ\!\Rightarrow}^{r,r-1}
\hat{\rho}_-^{r-1,\cdots,2},\label{rhohatAodd} \\
&\hat{\rho}(A^{(2)}_{2r})=\begin{cases}
\displaystyle \frac{2}{\hat{p}}\hat{\rho}^{1,0}_{\circ\!\Rightarrow}&\quad (r=1)\vspace{0.2cm} \\
\displaystyle \hat{\rho}_{\circ\!\Leftarrow}^{0,1}
\hat{\rho}_{-}^{1,\cdots,r-1}
\hat{\rho}_{\circ\!\Rightarrow}^{r,r-1}
\hat{\rho}_{-}^{r-1,\cdots,1}&\quad (r\ge 2)
\end{cases}
,\label{rhohatAeven}\\
&\hat{\rho}(D^{(2)}_{r+1})=\hat{\rho}_{\circ\!\Leftarrow}^{0,1}\hat{\rho}^{1,\cdots,r-1}_-\hat{\rho}^{r,r-1}_{\circ\!\Leftarrow}\hat{\rho}^{r-1,\cdots,1}_-,\label{rhohatD}
\end{align}
\end{subequations}
where
\begin{subequations}
\label{rhobb}
\begin{align}
&\hat{\rho}^{0,\cdots,s}_-=
(\hat{\mu}^{(0)})^{\frac{1}{2}}
\frac{1}{2\cosh\frac{\hat{p}}{2}}
\hat{\mu}^{(1)}
\frac{1}{2\cosh\frac{\hat{p}}{2}}
\cdots
\frac{1}{2\cosh\frac{\hat{p}}{2}}
\hat{\mu}^{(s-1)}
\frac{1}{2\cosh\frac{\hat{p}}{2}}
(\hat{\mu}^{(s)})^{\frac{1}{2}},\\
&\hat{\rho}_>^{0,1,2}=
(\mu^{(2)})^{\frac{1}{2}}\frac{1}{2\cosh\frac{\hat{p}}{2}}
\Bigl(
\hat{\mu}^{(0)}
\frac{\tanh\frac{\hat{p}}{2}}{2}
\hat{\mu}^{(1)}
+
\hat{\mu}^{(1)}
\frac{\tanh\frac{\hat{p}}{2}}{2}
\hat{\mu}^{(0)}
\Bigr)
\frac{1}{2\cosh\frac{\hat{p}}{2}}
(\mu^{(2)})^{\frac{1}{2}}
,\\
&\hat{\rho}_{\circ\!\Rightarrow}^{0,1}=
(\hat{\mu}^{(1)})^{\frac{1}{2}}
\Bigl(
\frac{\hat{p}}{4\pi\cosh\frac{\hat{p}}{2}}
\hat{\mu}^{(0)}
\frac{1}{2\cosh\frac{\hat{p}}{2}}
+
\frac{1}{2\cosh\frac{\hat{p}}{2}}
\hat{\mu}^{(0)}
\frac{\hat{p}}{4\pi\cosh\frac{\hat{p}}{2}}
\Bigr)
(\hat{\mu}^{(1)})^{\frac{1}{2}},\\
&\hat{\rho}_{\circ\!\Leftarrow}^{0,1}=
(\hat{\mu}^{(1)})^{\frac{1}{2}}
\frac{1}{2\cosh\frac{\hat{p}}{2}}
\hat{\mu}^{(0)}
\frac{2}{\hat{p}}
\hat{\mu}^{(0)}
\frac{1}{2\cosh\frac{\hat{p}}{2}}
(\hat{\mu}^{(1)})^{\frac{1}{2}},
\end{align}
\end{subequations}
with $\hat{\mu}^{(t)}=\mu^{(t)}(\hat{x})$.
The relation between the affine quivers and the density matrices can be summarized graphically as figure \ref{rhovsquiver}.
For simplicity in this section we consider the $D^{(2)}_{r+1}$ quiver matrix model only for even ranks, $Z(D^{(2)}_{r+1};2N)$.
The case with odd ranks, $Z(D^{(2)}_{r+1};2N-1)$, is treated in appendix \ref{app_sec_D(h)(2)r+1}.

\subsection{Building blocks}
\label{sec_buildingblocks}

In order to handle the matrix models for different affine quivers in a unified manner, we first notice that all matrix models are obtained by gluing some of the following building blocks by the convolution $\frac{1}{(2N)!}\int \frac{d^{2N}x}{(2\pi)^{2N}}$:
\begin{subequations}
\begin{align}
&\mbox{\raisebox{-0.9cm}{
\begin{tikzpicture}
\draw ([shift={(8.4,-6.3)}]-90:0.5) arc[radius=0.5, start angle=-90, end angle= 90];
\node at (8.4,-6.3) {$0$};
\draw (8.9,-6.3)--(9.1,-6.3);
\node at (9.5,-6.31) {$\cdots$};
\draw (9.9,-6.3)--(10.1,-6.3);
%
\draw ([shift={(10.6,-6.3)}]90:0.5) arc[radius=0.5, start angle=90, end angle=270];
\node at (10.6,-6.3) {$s$};
\draw [dashed] (8.4,-7.3)--(8.4,-5.3);
\draw [dashed] (10.6,-7.3)--(10.6,-5.3);
\end{tikzpicture}
}
}
\leftrightarrow {\cal O}_-^{0,\cdots,s}[x^{(0)}_i,x^{(s)}_i],\quad
\mbox{\raisebox{-1.1cm}{
\begin{tikzpicture}
\draw(8.4,-8.1) circle (0.5);
\node at (8.4,-8.1) {$0$};
\draw(8.4,-9.5) circle (0.5);
\node at (8.4,-9.5) {$1$};
\draw (8.85,-9.29)--(9.45,-9.01);
\draw (8.85,-8.31)--(9.45,-8.59);
\draw ([shift={(9.9,-8.8)}]90:0.5) arc[radius=0.5, start angle=90, end angle=270];
\node at (9.9,-8.8) {$2$};
%
%
\draw [dashed] (9.9,-9.8)--(9.9,-7.8);
\end{tikzpicture}
}}
\leftrightarrow {\cal O}_{>}^{0,1,2}[x^{(2)}_i],\\
&
\mbox{\raisebox{-0.9cm}{
\begin{tikzpicture}
\draw(8.4,-11) circle (0.5);
\node at (8.4,-11) {$0$};
\draw (8.89,-10.9)--(9.61,-10.9);
\draw (8.89,-11.1)--(9.61,-11.1);
\draw (9.13,-10.8)--(9.37,-11)--(9.13,-11.2);
\draw ([shift={(10.1,-11)}]90:0.5) arc[radius=0.5, start angle=90, end angle=270];
\node at (10.1,-11) {$1$};
\draw [dashed] (10.1,-12)--(10.1,-10);
\end{tikzpicture}
}
}
\leftrightarrow {\cal O}_{\circ\!\Rightarrow}^{0,1}[x^{(1)}_i],\quad
\mbox{\raisebox{-0.9cm}{
\begin{tikzpicture}
\draw(8.4,-13.2) circle (0.5);
\node at (8.4,-13.2) {$0$};
\draw (8.89,-13.1)--(9.61,-13.1);
\draw (8.89,-13.3)--(9.61,-13.3);
\draw (9.37,-13)--(9.13,-13.2)--(9.37,-13.4);
\draw ([shift={(10.1,-13.2)}]90:0.5) arc[radius=0.5, start angle=90, end angle=270];
\node at (10.1,-13.2) {$1$};
\draw [dashed] (10.1,-14.2)--(10.1,-12.2);
\end{tikzpicture}
}
}
\leftrightarrow {\cal O}_{\circ\!\Leftarrow}^{0,1}[x^{(1)}_i],
\end{align}
\end{subequations}
with
\begin{subequations}
\begin{align}
&{\cal O}_-^{0,\cdots,s}[x^{(0)}_i,x^{(s)}_i],\nonumber \\
&=
\prod_{i=1}^{2N}(\mu^{(0)}(x^{(0)}_i))^{\frac{1}{2}}
\prod_{i=1}^{2N}(\mu^{(s)}(x^{(s)}_i))^{\frac{1}{2}}
\prod_{t=1}^{s-1}\frac{1}{(2N)!}\int \prod_{i=1}^{2N}\frac{dx^{(t)}_i}{2\pi}\mu^{(t)}(x^{(t)}_i)\nonumber \\
&\quad\times \frac{
\prod_{i<j}^{2N}2\sinh\frac{x^{(0)}_i-x^{(0)}_j}{2}
\prod_{t=1}^{s-1}\prod_{i<j}^{2N}(2\sinh\frac{x^{(t)}_i-x^{(t)}_j}{2})^2
\prod_{i<j}^{2N}2\sinh\frac{x^{(s)}_i-x^{(s)}_j}{2}
}{
\prod_{t=0}^{s-1}\prod_{i,j}^{2N}2\cosh\frac{x^{(t)}-x^{(t+1)}}{2}
},\\
&{\cal O}_{>}^{0,1,2}[x^{(2)}_i]
=\prod_{i=1}^{2N}(\mu^{(2)}(x^{(2)}_i))^{\frac{1}{2}}\prod_{s=0}^1\frac{1}{N!}\int\prod_{i=1}^N\frac{dx^{(s)}_i}{2\pi}\mu^{(s)}(x^{(s)}_i)\nonumber \\
&\quad\quad\quad\quad\quad\quad\quad\quad\quad\times
\frac{\prod_{s=0}^1\prod_{i<j}^N(2\sinh\frac{x^{(s)}_i-x^{(s)}_j}{2})^2\prod_{i<j}^{2N}2\sinh\frac{x^{(2)}_i-x^{(2)}_j}{2}}{\prod_{s=0}^1\prod_{i=1}^N\prod_{j=1}^{2N}2\cosh\frac{x^{(s)}_i-x^{(2)}_j}{2}},\label{O>before} \\
&{\cal O}_{\circ\!\Rightarrow}^{0,1}[x^{(1)}_i]
=\prod_{i=1}^{2N}(\mu^{(1)}(x^{(1)}_i))^{\frac{1}{2}}\frac{1}{N!}\int\prod_{i=1}^N\frac{dx^{(0)}_i}{2\pi}\mu^{(0)}(x^{(0)}_i)\nonumber \\
&\quad\quad\quad\quad\quad\quad\quad\times
\frac{\prod_{i<j}^N(2\sinh\frac{x^{(0)}_i-x^{(0)}_j}{2})^4\prod_{i<j}^{2N}2\sinh\frac{x^{(1)}_i-x^{(1)}_j}{2}}{\prod_{i=1}^N\prod_{j=1}^{2N}(2\cosh\frac{x^{(0)}_i-x^{(1)}_j}{2})^2},\label{Oo=>before} \\
&{\cal O}_{\circ\!\Leftarrow}^{0,1}[x^{(1)}_i]
=\prod_{i=1}^{2N}(\mu^{(1)}(x^{(1)}_i))^{\frac{1}{2}}\frac{1}{(2N)!}\int\prod_{i=1}^{2N}\frac{dx^{(0)}_i}{2\pi}\mu^{(0)}(x^{(0)}_i)\nonumber \\
&\quad\quad\quad\quad\quad\quad\quad\times \frac{\prod_{i<j}^{2N}|2\sinh\frac{x^{(0)}_i-x^{(0)}_j}{2}|\prod_{i<j}^{2N}2\sinh\frac{x^{(1)}_i-x^{(1)}_j}{2}}{\prod_{i,j=1}^{2N}2\cosh\frac{x^{(0)}_i-x^{(1)}_j}{2}}.\label{Oo<=before}
\end{align}
\end{subequations}
See \eqref{ZBinbb}, \eqref{ZCinbb}, \eqref{ZAinbb} and \eqref{ZDinbb}.
Therefore let us first simplify these building blocks and re-express them in the one-dimensional quantum mechanical notation.
Before explaining the derivations, let us first summarize the final results:
\begin{subequations}
\label{Oall}
\begin{align}
&{\cal O}_{-}^{0,\cdots,s}[x^{(0)}_i,x^{(s)}_i]=\det\Bigl(\langle x^{(0)}_i|
\hat{\cal O}_{-}^{0,\cdots,s}
|x^{(s)}_j\rangle\Bigr),\label{O-} \\
&{\cal O}^{0,1,2}_>[x^{(2)}_i]
=(-1)^{\frac{N(N-1)}{2}}\frac{1}{N!}\int\prod_{i=1}^N\frac{dx^{(0)}_i}{2\pi}\det\begin{pmatrix}
[\langle x^{(0)}_i|\hat{\cal O}_{-}^{0,2}|x^{(2)}_j\rangle]_{i,j}^{N\times 2N}\vspace{0.2cm}\\
[\langle x^{(0)}_i|\hat{\cal O}_{>}^{0,1}\hat{O}^{1,2}_-|x^{(2)}_j\rangle]_{i,j}^{N\times 2N}
\end{pmatrix},\label{O>} \\
&{\cal O}^{0,1}_{\circ\!\Rightarrow}[x^{(1)}_i]=
(-1)^{\frac{N(N-1)}{2}}\frac{1}{N!}\int\frac{d^Nx^{(0)}_i}{(2\pi)^N}\det\begin{pmatrix}
[\langle x^{(0)}_i|\hat{\cal O}^{0,1}_{-}|x^{(1)}_j\rangle]_{i,j}^{N\times 2N}\vspace{0.2cm} \\
[\langle x^{(0)}_i|\hat{\cal O}^{0,1}_{\circ\!\Rightarrow}|x^{(1)}_j\rangle]_{i,j}^{N\times 2N}
\end{pmatrix},\label{Oo=>} \\
&{\cal O}^{0,1}_{\circ\!\Leftarrow}[x^{(1)}_i]=\frac{1}{(2N)!}\int\prod_{i=1}^{2N}\frac{dx^{(0)}_i}{2\pi}\det\Bigl(\langle x^{(0)}_i|(\hat{\mu}^{(0)})^{\frac{1}{2}}\hat{\cal O}^{0,1}_-|x^{(1)}_j\rangle\Bigr)\operatorname{pf}\Bigl(\langle x_i^{(0)}|\frac{-2i}{\hat{p}}|x^{(0)}_j\rangle\Bigr),\label{Oo<=}
\end{align}
\end{subequations}
with
\begin{subequations}
\label{Ohatall}
\begin{align}
&\hat{\cal O}_{-}^{0,\cdots,s}
=
(\hat{\mu}^{(0)})^{\frac{1}{2}}
\frac{1}{2\cosh\frac{\hat{p}}{2}}
\hat{\mu}^{(1)}
\frac{1}{2\cosh\frac{\hat{p}}{2}}
\hat{\mu}^{(2)}
\cdots
\hat{\mu}^{(s-1)}
\frac{1}{2\cosh\frac{\hat{p}}{2}}
(\hat{\mu}^{(s)})^{\frac{1}{2}},\\
&\hat{\cal O}^{0,1}_{>}=(\hat{\mu}^{(0)})^{\frac{1}{2}}\frac{-i\tanh\frac{\hat{p}}{2}}{2}(\hat{\mu}^{(1)})^{\frac{1}{2}},\\
&\hat{\cal O}^{0,1}_{\circ\!\Rightarrow}=(\hat{\mu}^{(0)})^{\frac{1}{2}}\frac{-i\hat{p}}{4\pi \cosh\frac{\hat{p}}{2}}(\hat{\mu}^{(1)})^{\frac{1}{2}}.
\end{align}
\end{subequations}
Note that the quantum mechanical operators $\hat{\cal O}^{0,\cdots,s}_{-}$, $\hat{\cal O}^{0,1}_{>}$ and $\hat{\cal O}^{0,1}_{\circ\!\Rightarrow}$ satisfy the following property under the transposition \eqref{Ot}
\begin{align}
(\hat{\cal O}^{0,\cdots,s}_{-})^{\rmT}
=\hat{\cal O}^{s,\cdots,0}_{-},\quad
(\hat{\cal O}^{0,1}_{>})^{\rmT}=-\hat{\cal O}^{1,0}_>,\quad
(\hat{\cal O}^{0,1}_{\circ\!\Rightarrow})^{\rmT}=-\hat{\cal O}^{1,0}_{\circ\!\Rightarrow}.
\end{align}
By using these properties, we can write ${\cal O}^{0,1}_{\circ\!\Rightarrow}[x^{(1)}_i]$ \eqref{Oo=>} and ${\cal O}^{0,1}_{\circ\!\Leftarrow}[x^{(1)}_i]$ \eqref{Oo<=} also as
\begin{align}
&{\cal O}^{0,1}_{\circ\!\Rightarrow}[x^{(1)}_i]
=(-1)^{\frac{N(N-1)}{2}}\frac{1}{N!}\int\frac{d^Nx^{(0)}_i}{(2\pi)^N}\det\Bigl(
[\langle x^{(1)}_i|\hat{\cal O}^{1,0}_{-}|x^{(0)}_j\rangle]_{i,j}^{2N\times N}\,\,\,\,
[-\langle x^{(1)}_i|\hat{\cal O}^{1,0}_{\circ\!\Rightarrow}|x^{(0)}_j\rangle]_{i,j}^{2N\times N}\Bigr),\\
&{\cal O}^{0,1}_{\circ\!\Leftarrow}[x^{(1)}_i]
=\frac{1}{(2N)!}\int\prod_{i=1}^{2N}\frac{dx^{(0)}_i}{2\pi}\det\Bigl(\langle x^{(1)}_i|\hat{\cal O}^{1,0}_-(\hat{\mu}^{(0)})^{\frac{1}{2}}|x^{(0)}_j\rangle\Bigr)\operatorname{pf}\Bigl(\langle x_i^{(0)}|\frac{-2i}{\hat{p}}|x^{(0)}_j\rangle\Bigr),
\end{align}
which are useful for later purpose.

Now let us derive \eqref{Oall} with \eqref{Ohatall}.
For ${\cal O}_{-}^{0,\cdots,s}[x^{(0)}_i,x^{(s)}_i]$, by using the Cauchy determinant formula \eqref{Cauchydetsinh} we find
\begin{align}
&{\cal O}_{-}^{0,\cdots,s}[x^{(0)}_i,x^{(s)}_i]\nonumber \\
&=
\prod_{i=1}^{2N}(\mu^{(0)}(x^{(0)}_i))^{\frac{1}{2}}
\prod_{i=1}^{2N}(\mu^{(s)}(x^{(s)}_i))^{\frac{1}{2}}
\prod_{t=1}^{s-1}\frac{1}{(2N)!}\int \prod_{i=1}^{2N}\frac{dx^{(t)}_i}{2\pi}\mu^{(t)}(x^{(t)}_i)
\prod_{t=0}^{s-1}
\det\langle x_i^{(t)}|\frac{1}{2\cosh\frac{\hat{p}}{2}}|x^{(t+1)}_j\rangle.
\end{align}
Then, by using the Andréief identity \eqref{Andreief} repeatedly we find \eqref{O-}.

For ${\cal O}_>^{0,1,2}[x^{(2)}_i]$, we first re-express the sinh factors for $x^{(0)}_i$ and $x^{(1)}_i$ in \eqref{O>before} as
\begin{align}
\prod_{s=0}^1\prod_{i<j}^N\Bigl(2\sinh\frac{x^{(s)}_i-x^{(s)}_j}{2}\Bigr)^2
=
\frac{
\prod_{i<j}^N2\sinh\frac{x^{(0)}_i-x^{(0)}_j}{2}
\prod_{i<j}^N2\sinh\frac{x^{(1)}_i-x^{(1)}_j}{2}
}{
\prod_{i,j}^N2\sinh\frac{x^{(0)}_i-x^{(1)}_j}{2}
}
\prod_{i<j}^{2N}2\sinh\frac{\tilde{x}^{(0)}_i-\tilde{x}_j^{(0)}}{2},
\label{sinhdecompose}
\end{align}
where $(\tilde{x}^{(0)}_j)_{j=1,\cdots,2N}=(
\tilde{x}^{(0)}_1,
\cdots,
\tilde{x}^{(0)}_N,
\tilde{x}^{(1)}_1,
\cdots,
\tilde{x}^{(1)}_N
)$
so that we can apply the Cauchy determinant formulas \eqref{Cauchydetsinh} and \eqref{Cauchydetsinhsinh} to the first factors in \eqref{sinhdecompose} and the remaining factors in ${\cal O}^{0,1,2}_>[x^{(2)}_i]$ separately.
Using the notation of one-dimensional quantum mechanics, we obtain
\begin{align}
{\cal O}^{0,1,2}_>[x^{(2)}_i]
&=
(-1)^{\frac{N(N-1)}{2}}
\prod_{i=1}^{2N}(\mu^{(2)}(x^{(2)}_i))^{\frac{1}{2}}
\prod_{s=0}^1\frac{1}{N!}\int\frac{dx^{(s)}_i}{2\pi}\mu^{(s)}(x^{(s)}_i)\nonumber \\
&\quad\times
\det\Bigl(\langle x^{(0)}_i|\Bigl(-\frac{i\tanh\frac{\hat{p}}{2}}{2}\Bigr)|x^{(1)}_j\rangle\Bigr)
\det\begin{pmatrix}
[\langle x^{(0)}_i|\frac{1}{2\cosh\frac{\hat{p}}{2}}|x^{(2)}_j\rangle]_{i,j}^{N\times 2N}\vspace{0.2cm} \\
[\langle x^{(1)}_i|\frac{1}{2\cosh\frac{\hat{p}}{2}}|x^{(2)}_j\rangle]_{i,j}^{N\times 2N}
\end{pmatrix}.
\end{align}
By applying the Andréief identity \eqref{Andreief} to the $x^{(1)}_i$-integrations, we obtain \eqref{O>}.

For ${\cal O}^{0,1}_{\circ\!\Rightarrow}[x^{(1)}_i]$, we first rewrite the factors involving $x^{(0)}_i$ in \eqref{Oo=>before} as
\begin{subequations}
\begin{align}
\prod_{i<j}^N\Bigl(2\sinh\frac{x^{(0)}_i-x^{(0)}_j}{2}\Bigr)^4&=\lim_{y^{(0)}_i\rightarrow x^{(0)}_i}(-1)^{\frac{N(N-1)}{2}}\frac{
\prod_{i<j}^{2N}2\sinh\frac{\tilde{x}^{(0)}_i-\tilde{x}^{(0)}_j}{2}
}{
\prod_{i=1}^N 2\sinh\frac{x^{(0)}_i-y^{(0)}_i}{2}
},\\
\frac{1}{\prod_{i=1}^N\prod_{j=1}^{2N}(2\cosh\frac{x^{(0)}_i-x^{(1)}_j}{2})^2}
&=\lim_{y_i^{(0)}\rightarrow x_i^{(0)}}\frac{1}{\prod_{i,j=1}^{2N}2\cosh\frac{\tilde{x}^{(0)}_i-x^{(1)}_j}{2}}
\end{align}
\end{subequations}
with $(\tilde{x}^{(0)}_i)_{i=1,\cdots,2N}=(x^{(0)}_1,\cdots,x^{(0)}_N,y^{(0)}_1,\cdots,y^{(0)}_N)$.
Plugging these into ${\cal O}^{0,1}_{\circ\!\Rightarrow}[x^{(1)}_i]$ and using the Cauchy determinant formula \eqref{Cauchydetsinh} for the $\tilde{x}^{(0)}_i,x^{(1)}_i$-products, we obtain
\begin{align}
{\cal O}^{0,1}_{\circ\!\Rightarrow}[x^{(1)}_i]&=
(-1)^{\frac{N(N-1)}{2}}\prod_{i=1}^{2N}(\mu^{(1)}(x^{(1)}_i))^{\frac{1}{2}}\frac{1}{N!}\int \prod_{i=1}^N\frac{dx_i^{(0)}}{2\pi}\mu^{(0)}(x^{(0)}_i)\nonumber \\
&\quad\times \lim_{y^{(0)}_i\rightarrow x^{(0)}_i}
\frac{1}{
\prod_{i=1}^N 2\sinh\frac{x^{(0)}_i-y^{(0)}_i}{2}
}
\det\begin{pmatrix}
\displaystyle \left[\frac{1}{2\cosh\frac{x^{(0)}_i-x^{(1)}_j}{2}}\right]_{i,j}^{N\times 2N}\vspace{0.2cm} \\
\displaystyle \left[\frac{1}{2\cosh\frac{y^{(0)}_i-x^{(1)}_j}{2}}\right]_{i,j}^{N\times 2N}
\end{pmatrix}
\end{align}
The limit can be taken by performing the elementary row transformation as
\begin{align}
&\lim_{y^{(0)}_i\rightarrow x^{(0)}_i}
\frac{1}{\prod_{i=1}^N2\sinh\frac{x^{(0)}_i-y^{(0)}_i}{2}}\det\begin{pmatrix}
\displaystyle\left[\frac{1}{2\cosh\frac{x^{(0)}_i-x^{(1)}_j}{2}}\right]_{i,j}^{N\times 2N}\vspace{0.2cm} \\
\displaystyle\left[\frac{1}{2\cosh\frac{y^{(0)}_i-x^{(1)}_j}{2}}\right]_{i,j}^{N\times 2N}
\end{pmatrix}\nonumber \\
&=
\det\begin{pmatrix}
\displaystyle \left[\frac{1}{2\cosh\frac{x^{(0)}_i-x^{(1)}_j}{2}}\right]_{i,j}^{N\times 2N}\vspace{0.2cm} \\
\displaystyle \left[\lim_{y^{(0)}_i\rightarrow x^{(0)}_i}
\frac{1}{
2\sinh\frac{x^{(0)}_i-y^{(0)}_i}{2}
}
\left(
\frac{1}{2\cosh\frac{y^{(0)}_i-x^{(1)}_j}{2}}
-\frac{1}{2\cosh\frac{x^{(0)}_i-x^{(1)}_j}{2}}
\right)\right]_{i,j}^{N\times 2N}
\end{pmatrix}\nonumber \\
&=
\det\begin{pmatrix}
\displaystyle\left[\frac{1}{2\cosh\frac{x^{(0)}_i-x^{(1)}_j}{2}}\right]_{i,j}^{N\times 2N}\vspace{0.2cm} \\
\displaystyle\left[\frac{
\sinh\frac{x^{(0)}_i-x^{(1)}_i}{2}
}{
(
2\cosh\frac{x^{(0)}_i-x^{(1)}_j}{2})^2
}\right]_{i,j}^{N\times 2N}
\end{pmatrix}
\end{align}
Plugging this into \eqref{Oo=>before} and using the notation of one-dimensional quantum mechanics, we finally obtain \eqref{Oo=>}.

For ${\cal O}^{0,1}_{\circ\!\Leftarrow}[x^{(1)}_i]$, we first rewrite the sinh factors of $x^{(0)}_i$ by using the formula \eqref{VdMlikeforsgn} as
\begin{align}
\prod_{i<j}^{2N}\Bigl|2\sinh\frac{x^{(0)}_i-x^{(0)}_j}{2}\Bigr|
=\prod_{i<j}^{2N}2\sinh\frac{x^{(0)}_i-x^{(0)}_j}{2}
\operatorname{pf}\Bigl(\langle x^{(0)}_i|\frac{-2i}{\hat{p}}|x^{(0)}_j\rangle\Bigr).
\end{align}
Plugging this into \eqref{Oo<=before} and using the Cauchy determinant formula \eqref{Cauchydetsinh} to the sinh factor without absolute value symbols, we obtain \eqref{Oo<=}.

\subsection{$B^{(1)}_r$}

Now let us derive the Fredholm determinant expression \eqref{Xi=sqrtDet} for the quiver matrix models by using the building blocks \eqref{Oall}.
For $\Gamma=B^{(1)}_r$ the quiver matrix models can be written as
\begin{align}
&Z(B^{(1)}_r;N)\nonumber \\
&=
\frac{1}{((2N)!)^2}
\int\prod_{i=1}^{2N}\frac{dx^{(2)}_i}{2\pi}
\prod_{i=1}^{2N}\frac{dx^{(r-1)}_i}{2\pi}
{\cal O}_{>}^{0,1,2}[x^{(2)}_i]
{\cal O}_{-}^{2,\cdots,r-1}[x^{(2)}_i,x^{(r-1)}_i]
{\cal O}_{\circ\!\Leftarrow}^{r,r-1}[x^{(r-1)}_i].
\label{ZBinbb}
\end{align}
Substituting the building blocks
and performing the $x^{(2)}_i$,$x^{(r-1)}_i$-integrations by using the Andr\'eief identity \eqref{Andreief}, we obtain
\begin{align}
&Z(B^{(1)}_r;N)\nonumber \\
&=\frac{(-1)^{\frac{N(N-1)}{2}}}{N!(2N)!}\int \prod_{i=1}^N\frac{dx^{(0)}_i}{2\pi}\prod_{i=1}^{2N}\frac{dx^{(r)}_i}{2\pi}
\det\begin{pmatrix}
[\langle x^{(0)}_i|\hat{\cal O}^{0,2,\cdots,r}_-(\hat{\mu}^{(r)})^{\frac{1}{2}}|x^{(r)}_j\rangle]_{i,j}^{N\times 2N}\vspace{0.2cm} \\
[\langle x^{(0)}_i|\hat{\cal O}^{0,1}_>\hat{\cal O}^{1,2,\cdots,r}_-(\hat{\mu}^{(r)})^{\frac{1}{2}}|x^{(r)}_j\rangle]_{i,j}^{N\times 2N}
\end{pmatrix}\nonumber \\
&\quad\times
\operatorname{pf}\Bigl(\langle x^{(r)}_i|\frac{-2i}{\hat{p}}|x^{(r)}_j\rangle\Bigr).
\end{align}
By further performing the $x^{(r)}_i$-integrations by using De Bruijn's formula \eqref{Bruijn}, we obtain
\begin{align}
&Z(B^{(1)}_r;N)\nonumber \\
&=\frac{(-1)^{\frac{N(N-1)}{2}}}{N!}\int \prod_{i=1}^N\frac{dx_i}{2\pi}\nonumber \\
&\quad\times\operatorname{pf}\begin{pmatrix}
[\langle x_i|\hat{H}_B|x_j\rangle]_{i,j}^{N\times N}
&
[-\langle x_i|\hat{H}_B(\frac{\hat{\mu}^{(1)}}{\hat{\mu}^{(0)}})^{\frac{1}{2}}\hat{\cal O}^{1,0}_>|x_j\rangle]_{i,j}^{N\times N}\vspace{0.2cm} \\
[\langle x_i|\hat{\cal O}^{0,1}_>(\frac{\hat{\mu}^{(1)}}{\hat{\mu}^{(0)}})^{\frac{1}{2}}\hat{H}_B|x_j\rangle]_{i,j}^{N\times N}
&
[-\langle x_i|\hat{\cal O}^{0,1}_>(\frac{\hat{\mu}^{(1)}}{\hat{\mu}^{(0)}})^{\frac{1}{2}}\hat{H}_B(\frac{\hat{\mu}^{(1)}}{\hat{\mu}^{(0)}})^{\frac{1}{2}}\hat{\cal O}^{1,0}_>|x_j\rangle]_{i,j}^{N\times N}
\end{pmatrix},
\end{align}
with
\begin{align}
\hat{H}_B=
\hat{\cal O}^{0,2,\cdots,r}_-
(\hat{\mu}^{(r)})^{\frac{1}{2}}
\frac{-2i}{\hat{p}}
(\hat{\mu}^{(r)})^{\frac{1}{2}}
\hat{\cal O}^{r,\cdots,2,0}_-.
\end{align}
Hence the grand canonical sum \eqref{GC} of the matrix model is rewritten by using the Fredholm Pfaffian formula \eqref{FredholmPfaffian} as
\begin{align}
&\Xi(B^{(1)}_r;u)=\sqrt{\operatorname{Det}\begin{pmatrix}
u\hat{H}_B
&
1-u\hat{H}_B(\frac{\mu^{(1)}}{\mu^{(0)}})^{\frac{1}{2}}\hat{\cal O}^{1,0}_>\\
-1+u\hat{\cal O}^{0,1}_>(\frac{\mu^{(1)}}{\mu^{(0)}})^{\frac{1}{2}}\hat{H}_B
&
-u\hat{\cal O}^{0,1}_>(\frac{\mu^{(1)}}{\mu^{(0)}})^{\frac{1}{2}}\hat{H}_B(\frac{\mu^{(1)}}{\mu^{(0)}})^{\frac{1}{2}}\hat{\cal O}^{1,0}_>
\end{pmatrix}
}.
\end{align}
By performing the elementary row transformations, we obtain
\begin{align}
\Xi(B^{(1)}_r;u)&=\sqrt{\operatorname{Det}\begin{pmatrix}
0
&
1+u(
-\hat{H}_B(\frac{\mu^{(1)}}{\mu^{(0)}})^{\frac{1}{2}}\hat{\cal O}^{1,0}_>
-\hat{H}_B\hat{\cal O}^{0,1}_>(\frac{\mu^{(1)}}{\mu^{(0)}})^{\frac{1}{2}})\\
-1&0
\end{pmatrix}
}\nonumber \\
&=\sqrt{\operatorname{Det}(1+u\hat{\rho}(B^{(1)}_r))}
\end{align}
where $\hat{\rho}(B^{(1)}_r)$ is
\begin{align}
\hat{\rho}(B^{(1)}_r)
=- \hat{H}_B\Bigl(
\Bigl(\frac{\hat{\mu}^{(1)}}{\hat{\mu}^{(0)}}\Bigr)^{\frac{1}{2}}\hat{\cal O}^{1,0}_>
+\hat{\cal O}^{0,1}_>\Bigl(\frac{\hat{\mu}^{(1)}}{\hat{\mu}^{(0)}}\Bigr)^{\frac{1}{2}}
\Bigr),
\end{align}
which coincides with \eqref{rhohatB} up to the similarity transformation.

\subsection{$C^{(1)}_r$}

For $\Gamma=C^{(1)}_r$, the quiver matrix model can be written as
\begin{align}
&Z(C^{(1)}_r;N)\nonumber \\
&= \frac{1}{((2N)!)^2}
\int\prod_{i=1}^{2N}\frac{dx^{(1)}_i}{2\pi}
\prod_{i=1}^{2N}\frac{dx^{(r-1)}_i}{2\pi}
{\cal O}_{\circ\!\Rightarrow}^{0,1}[x^{(1)}_i]
{\cal O}_{-}^{1,\cdots,r-1}[x^{(1)}_i,x^{(r-1)}_i]
{\cal O}_{\circ\!\Leftarrow}^{r,r-1}[x^{(r-1)}_i].
\label{ZCinbb}
\end{align}
Substituting the building blocks and performing the $x^{(1)}_i$,$x^{(r-1)}_i$-integrations by using the Andr\'eief identity \eqref{Andreief}, we obtain
\begin{align}
&Z(C^{(1)}_r;N)\nonumber \\
&=\frac{1}{(N!)^2}\int \prod_{i=1}^N\frac{dx^{(0)}_i}{2\pi}\prod_{i=1}^{N}\frac{dx^{(r)}_i}{2\pi}\nonumber \\
&\quad\times
\det\begin{pmatrix}
[\langle x^{(0)}_i|\hat{\cal O}^{0,\cdots,r}_-|x^{(r)}_j\rangle]_{i,j}^{N\times N}
&
[-\langle x^{(0)}_i|\hat{\cal O}^{0,\cdots,r-1}_-\hat{\cal O}^{r-1,r}_{\circ\!\Rightarrow}|x^{(r)}_j\rangle]_{i,j}^{N\times N}\vspace{0.2cm} \\
[\langle x^{(0)}_i| \hat{\cal O}^{0,1}_{\circ\!\Rightarrow} \hat{\cal O}^{1,\cdots,r}_-|x^{(r)}_j\rangle]_{i,j}^{N\times N}
&
[-\langle x^{(0)}_i|\hat{\cal O}^{0,1}_{\circ\!\Rightarrow}\hat{\cal O}^{1,\cdots,r-1}_-\hat{\cal O}^{r-1,r}_{\circ\!\Rightarrow}|x^{(r)}_j\rangle]_{i,j}^{N\times N}
\end{pmatrix}.
\end{align}
By performing the $x^{(r)}_i$-integrations by using the Pfaffian analogue of the Andréief identity \eqref{CBPfaffian}, we obtain
\begin{align}
&Z(C^{(1)}_r;N)\nonumber \\
&=\frac{(-1)^{\frac{N(N-1)}{2}}}{N!}\int \prod_{i=1}^N\frac{dx_i}{2\pi}\nonumber \\
&\quad\times
\operatorname{pf}\begin{pmatrix}
[\langle x_i|\hat{H}_C|x_j\rangle]_{i,j}^{N\times N}
&
[-\langle x_i|\hat{H}_C(\hat{\cal O}^{1,0}_-)^{-1}\hat{\cal O}^{1,0}_{\circ\!\Rightarrow}|x_j\rangle]_{i,j}^{N\times N}\vspace{0.2cm} \\
[\langle x_i| \hat{\cal O}^{0,1}_{\circ\!\Rightarrow} (\hat{\cal O}^{0,1}_-)^{-1} \hat{H}_C |x_j\rangle]_{i,j}^{N\times N}
&
[-\langle x_i| \hat{\cal O}^{0,1}_{\circ\!\Rightarrow} (\hat{\cal O}^{0,1}_-)^{-1} \hat{H}_C 
(\hat{\cal O}^{1,0}_-)^{-1}
\hat{\cal O}^{1,0}_{\circ\!\Rightarrow}
|x_j\rangle]_{i,j}^{N\times N}
\end{pmatrix}
\end{align}
with
\begin{align}
\hat{H}_C=
\hat{\cal O}^{0,\cdots,r}_-
\hat{\cal O}^{r,r-1}_{\circ\!\Rightarrow}
\hat{\cal O}^{r-1,\cdots,0}_-
+
\hat{\cal O}^{0,\cdots,r-1}_-
\hat{\cal O}^{r-1,r}_{\circ\!\Rightarrow}
\hat{\cal O}^{r,\cdots,0}_-.
\end{align}
Hence the grand canonical sum \eqref{GC} of the matrix model is rewritten by using the Fredholm Pfaffian formula \eqref{FredholmPfaffian} as
\begin{align}
&\Xi(C^{(1)}_r;u)=\sqrt{\operatorname{Det}\begin{pmatrix}
u\hat{H}_C
&
1-u\hat{H}_C(\hat{\cal O}^{1,0}_-)^{-1}\hat{\cal O}^{1,0}_{\circ\!\Rightarrow}\\
-1+u \hat{\cal O}^{0,1}_{\circ\!\Rightarrow} (\hat{\cal O}^{0,1}_-)^{-1} \hat{H}_C
&
-u \hat{\cal O}^{0,1}_{\circ\!\Rightarrow} (\hat{\cal O}^{0,1}_-)^{-1} \hat{H}_C(\hat{\cal O}^{1,0}_-)^{-1}\hat{\cal O}^{1,0}_{\circ\!\Rightarrow}
\end{pmatrix}
}\nonumber \\
&=\sqrt{\operatorname{Det}(1+u\hat{\rho}(C^{(1)}_r))},
\end{align}
where $\hat{\rho}(C^{(1)}_r)$ is
\begin{align}
\hat{\rho}(C^{(1)}_r)
=-\hat{H}_C
(
(\hat{\cal O}^{1,0}_-)^{-1}
\hat{\cal O}^{1,0}_{\circ\!\Rightarrow}
+
\hat{\cal O}^{0,1}_{\circ\!\Rightarrow}
(\hat{\cal O}^{0,1}_-)^{-1}
),
\end{align}
which coincides with \eqref{rhohatC} up to a similarity transformation.

\subsection{$A^{(2)}_{2r-1}$ and $A^{(2)}_{2r}$}

Next let us consider the quiver matrix models with $\Gamma=A^{(2)}_{2r-1}$ and $\Gamma=A^{(2)}_{2r}$, which can be written as
\begin{subequations}
\label{ZAinbb}
\begin{align}
&Z(A^{(2)}_{2r-1};N)\nonumber \\
&= \frac{1}{((2N)!)^2}
\int\prod_{i=1}^{2N}\frac{dx^{(2)}_i}{2\pi}
\prod_{i=1}^{2N}\frac{dx^{(r-1)}_i}{2\pi}
{\cal O}_{>}^{0,1,2}[x^{(2)}_i]
{\cal O}_{-}^{2,\cdots,r-1}[x^{(2)}_i,x^{(r-1)}_i]
{\cal O}_{\circ\!\Rightarrow}^{r,r-1}[x^{(r-1)}_i],\\
&Z(A^{(2)}_{2r};N)\nonumber \\
&= \frac{1}{((2N)!)^2}
\int\prod_{i=1}^{2N}\frac{dx^{(1)}_i}{2\pi}
\prod_{i=1}^{2N}\frac{dx^{(r-1)}_i}{2\pi}
{\cal O}_{\circ\Leftarrow}^{0,1}[x^{(1)}_i]
{\cal O}_{-}^{1,\cdots,r-1}[x^{(1)}_i,x^{(r-1)}_i]
{\cal O}_{\circ\!\Rightarrow}^{r,r-1}[x^{(r-1)}_i].
\end{align}
\end{subequations}
The derivations of the Fermi gas formalism for these models are parallel to those for $C^{(1)}_r$ and $B^{(1)}_r$ respectively.
Namely, for $A^{(2)}_{2r-1}$, substituting the building blocks, performing the $x^{(2)}_i,x^{(r-1)}_i$-integrations by the Andréief identity \eqref{Andreief}, performing the $x^{(r)}_i$-integrations by the Pfaffian formula \eqref{CBPfaffian} and then using the Fredholm Pfaffian formula \eqref{FredholmPfaffian}, we finally obtain $\sum_{N=0}^\infty u^N Z(A^{(2)}_{2r-1};N)=\sqrt{\operatorname{Det}(1+u\hat{\rho}(A^{(2)}_{2r-1}))}$ with $\hat{\rho}(A^{(2)}_{2r-1})$ given by \eqref{rhohatAodd} up to a similarity transformation.
For $A^{(2)}_{2r-1}$, substituting the building blocks, performing the $x^{(1)}_i,x^{(r-1)}_i$-integrations by the Andréief identity \eqref{Andreief}, performing the $x^{(0)}_i$-integrations by the De Bruijn's formula \eqref{Bruijn} and then using the Fredholm Pfaffian formula \eqref{FredholmPfaffian}, we finally obtain $\sum_{N=0}^\infty u^N Z(A^{(2)}_{2r};N)=\sqrt{\operatorname{Det}(1+u\hat{\rho}(A^{(2)}_{2r}))}$ with $\hat{\rho}(A^{(2)}_{2r})$ given by \eqref{rhohatAeven} up to a similarity transformation.

\subsection{$D^{(2)}_{r+1}$ with even ranks}

The quiver matrix model for $\Gamma=D^{(2)}_{r+1}$ can be written as
\begin{align}
&Z(D^{(2)}_{r+1};2N)\nonumber \\
&= \frac{1}{((2N)!)^2}
\int\prod_{i=1}^{2N}\frac{dx^{(1)}_i}{2\pi}
\prod_{i=1}^{2N}\frac{dx^{(r-1)}_i}{2\pi}
{\cal O}_{\circ\Leftarrow}^{0,1}[x^{(1)}_i]
{\cal O}_{-}^{1,\cdots,r-1}[x^{(1)}_i,x^{(r-1)}_i]
{\cal O}_{\circ\!\Leftarrow}^{r,r-1}[x^{(r-1)}_i].
\label{ZDinbb}
\end{align}
Substituting the building blocks and performing the $x^{(1)}_i,x^{(r-1)}_i$-integration by Andréief identity \eqref{Andreief} and then performing $x^{(r)}_i$ integrations by De Bruijn's formula \eqref{Bruijn}, we obtain
\begin{align}
Z(D^{(2)}_{r+1};2N)&
=\frac{1}{(2N)!}\int\prod_{i=1}^{2N}\frac{dx^{(0)}_i}{2\pi}\operatorname*{pf}_{2N\times 2N}
\Bigl(
\langle x^{(0)}_i|\hat{H}_D|x^{(0)}_j\rangle\Bigr)
\operatorname*{pf}_{2N\times 2N}\Bigl(\langle x^{(0)}_i|\frac{-2i}{\hat{p}}|x^{(0)}_j\rangle\Bigr)\nonumber \\
&=\frac{1}{(2N)!}\int\prod_{i=1}^{2N}\frac{dx^{(0)}_i}{2\pi}
\operatorname*{pf}_{4N\times 4N}
\begin{pmatrix}
[\langle x^{(0)}_i|\hat{H}_D|x^{(0)}_j\rangle]_{i,j}^{2N\times 2N}&[0]^{2N\times 2N}\vspace{0.2cm} \\
[0]^{2N\times 2N}&[\langle x^{(0)}_i|\frac{-2i}{\hat{p}}|x^{(0)}_j\rangle]_{i,j}^{2N\times 2N}
\end{pmatrix},
\end{align}
where
\begin{align}
\hat{H}_D=
(\hat{\mu}^{(0)})^{\frac{1}{2}}
\hat{\cal O}^{0,\cdots,r}_-
(\hat{\mu}^{(r)})^{\frac{1}{2}}
\frac{-2i}{\hat{p}}
(\hat{\mu}^{(r)})^{\frac{1}{2}}
\hat{\cal O}^{0,\cdots,r}_-
(\hat{\mu}^{(0)})^{\frac{1}{2}}.
\end{align}
Now let us consider the grand canonical sum of $Z(D^{(2)}_{r+1};2N)$
\begin{align}
\Xi(D^{(2)}_{r+1};u)
&=\sum_{N=0}^\infty u^NZ(D^{(2)}_{r+1};2N)\nonumber \\
&=\sum_{M\in 2\mathbb{Z}_{\ge 0}}^\infty (-u)^{\frac{M}{2}}\frac{(-1)^{\frac{M(M-1)}{2}}}{M!}
\int\prod_{i=1}^{M}\frac{dx^{(0)}_i}{2\pi}\nonumber \\
&\quad\times
\operatorname*{pf}_{2M\times 2M}
\begin{pmatrix}
[\langle x^{(0)}_i|\hat{H}_D|x^{(0)}_j\rangle]_{i,j}^{M\times M}&[0]^{M\times M}\vspace{0.2cm} \\
[0]^{M\times M}&[\langle x^{(0)}_i|\frac{-2i}{\hat{p}}|x^{(0)}_j\rangle]_{i,j}^{M\times M}
\end{pmatrix}.
\end{align}
Here in the second expression we have denoted $2N=M$.
Since the determinant of an $M\times M$ antisymmetric matrix vanishes trivially for odd $M$, we can extend the range of summation from $M\in 2\mathbb{Z}_{\ge 0}$ to $M\in\mathbb{Z}_{\ge 0}$.
After that we can apply the Fredholm Pfaffian formula \eqref{FredholmPfaffian} as
\begin{align}
\Xi(D^{(2)}_{r+1};u)=\sqrt{\operatorname{Det}\begin{pmatrix}
(-u)^{\frac{1}{2}}\hat{H}_D&1\\
-1&(-u)^{\frac{1}{2}}(\frac{-2i}{\hat{p}})
\end{pmatrix}}.
\end{align}
By performing elementary row transformation we obtain
\begin{align}
\Xi(D^{(2)}_{r+1};u)=\sqrt{\operatorname{Det}(1+u\hat{\rho}(D^{(2)}_{r+1}))}
\end{align}
with
\begin{align}
\hat{\rho}^{(2)}_{r+1}=-\hat{H}_D\frac{-2i}{\hat{p}},
\end{align}
which coincides with \eqref{rhohatD} up to a similarity transformation.

\section{Large $N$ free energy via Fermi gas formalism}
\label{sec_N32byFermigas}

In this section we calculate, by using the Fermi gas formalism \eqref{Xi=sqrtDet}, the leading behavior of the free energy $-\log Z(\Gamma;N)$ of the matrix model in the large $N$ limit when the measure factors $\mu^{(s)}(x)$ are given by the Fresnel measures $\mu^{(s)}(x)=e^{\frac{ik_s}{4\pi}x^2}$ with $k_s$ satisfying the level balance condition \eqref{level_balance}.

First we notice that the relation between the matrix model and the grand canonical partition function can be inverted by
\begin{align}
Z(\Gamma;N)=\int^{i\infty}_{-i\infty}\frac{d\mu}{2\pi i}e^{\frac{1}{2}J(\mu)-\mu N}
\label{inversetrsf}
\end{align}
where $J(\mu)$ is the modified grand potential defined by
\begin{align}
\sum_{n=-\infty}^\infty e^{\frac{1}{2}J(\mu+2\pi in)}=\sqrt{\operatorname*{Det}(1+e^\mu\hat{\rho}(\Gamma))}.
\end{align}
The goal of our calculation below is to obtain the large $\mu$ expansion of $J(\mu)$ as
\begin{align}
J(\mu)=\frac{2C(\Gamma)}{3}\mu^3+\cdots,
\label{Jinlargemu}
\end{align}
with $C(\Gamma)$ a model dependent constant.
See \eqref{CofB(h)(1)_r}, \eqref{CofC(h)(1)_r}, \eqref{CofA(h)(2)_2r-1}, \eqref{CofA(h)(2)_2r} and \eqref{CofD(h)(2)_r+1} for its explicit expression.
Plugging this into the inverse transformation \eqref{inversetrsf} and approximating the integration by the contribution from the saddle ponit $\mu_*=\sqrt{\frac{N}{C(\Gamma)}}$, we obtain
\begin{align}
-\log Z(\Gamma;N)=\frac{2}{3\sqrt{C(\Gamma)}}N^{\frac{3}{2}}+\cdots.
\label{largeNfreeenergyfromFG}
\end{align}

Since the calculation is similar for all the quivers considered in the previous section, here let us focus only on the quiver $B^{(1)}_r$.
Note that our calculation is also parallel to that for the standard circular quiver Chern--Simons theories performed in \cite{Marino:2011eh} except the sub-leading corrections in $1/N$ which we mention in the end of the section.
For the $B^{(1)}_r$ quiver, the level balance condition is
\begin{align}
k_0+k_1+2\sum_{s=2}^rk_s=0,
\end{align}
whose general solution can be parametrized, as explained in section~\ref{sec:aff_quiver_matrix_model}, in terms of the simple roots, $\alpha_0 = - e_1 - e_2$, $\alpha_s = e_s - e_{s+1}$ for $s = 1,\ldots,r-1$, and $\alpha_r = e_r$.
Then, we have
\begin{subequations}
\begin{align}
&k_0=k(-\sigma_0-\sigma_1),\quad
k_1=k(\sigma_0-\sigma_1),\\
&k_s=k(\sigma_{s-1}-\sigma_s),\quad (s=2,\cdots,r-1),\\
&k_r=k\sigma_{r-1},
\end{align}
\end{subequations}
where we put an overall factor $k$ for latter convenience.
Substituting $\mu^{(s)}(x)=e^{\frac{ik_s}{4\pi}x^2}$ with these levels $k_s$ into the density matrix $\hat{\rho}(B^{(1)}_r)$ \eqref{rhohatB} with \eqref{rhobb}, we obtain, up to a similarity transformation,
\begin{align}
\hat{\rho}(B^{(1)}_r)&=
\Bigl(
\tanh\frac{\hat{p}-\sigma_0\hat{x}'}{2}
+\tanh\frac{\hat{p}+\sigma_0\hat{x}'}{2}
\Bigr)
\frac{1}{2\cosh\frac{\hat{p}+\sigma_1\hat{x}'}{2}}
\frac{1}{2\cosh\frac{\hat{p}+\sigma_2\hat{x}'}{2}}
\cdots
\frac{1}{2\cosh\frac{\hat{p}+\sigma_{r-1}\hat{x}'}{2}}
\frac{1}{\hat{p}}\nonumber \\
&\quad
\frac{1}{2\cosh\frac{\hat{p}-\sigma_{r-1}\hat{x}'}{2}}
\frac{1}{2\cosh\frac{\hat{p}-\sigma_{r-2}\hat{x}'}{2}}
\cdots
\frac{1}{2\cosh\frac{\hat{p}-\sigma_1\hat{x}'}{2}},
\label{rhohatBCS}
\end{align}
where we have rescaled the position operator as $\hat{x}'=k\hat{x}$ so that it satisfies the canonical commutation relation
\begin{align}
[\hat{x}',\hat{p}]=i\hbar,\quad \hbar=2\pi k.
\end{align}

Now let us consider the small $k$ expansion, which corresponds to the semiclassical expansion of the Fermi gas system.
In this limit we can calculate the modified grand potential $J(\mu)$ as \cite{Marino:2011eh}
\begin{align}
J(\mu)=\text{Tr}\log (1+e^\mu\hat{\rho}(B^{(1)}_r))
=\int_0^\infty dE\frac{dn}{dE}\log(1+e^{\mu-E}),
\label{Jfromn(E)}
\end{align}
where $n(E)$ is the number of states
\begin{align}
n(E)=\operatorname{Tr}\theta(\hat{H}-E),\quad \theta(z)=
\begin{cases}
1&\quad (z\ge 0)\\
0&\quad (z<0)
\end{cases}
\end{align}
with $\hat{H}$ given by $\hat{\rho}(B^{(1)}_r)=e^{-\hat{H}}$.
Note that in the rewriting of $J(\mu)$ in terms of $n(E)$, we have assumed that the energy spectrum is discrete, bounded from below $E\ge 0$ and that $n(0)=0$.

\begin{figure}
\begin{center}
\includegraphics[width=8cm]{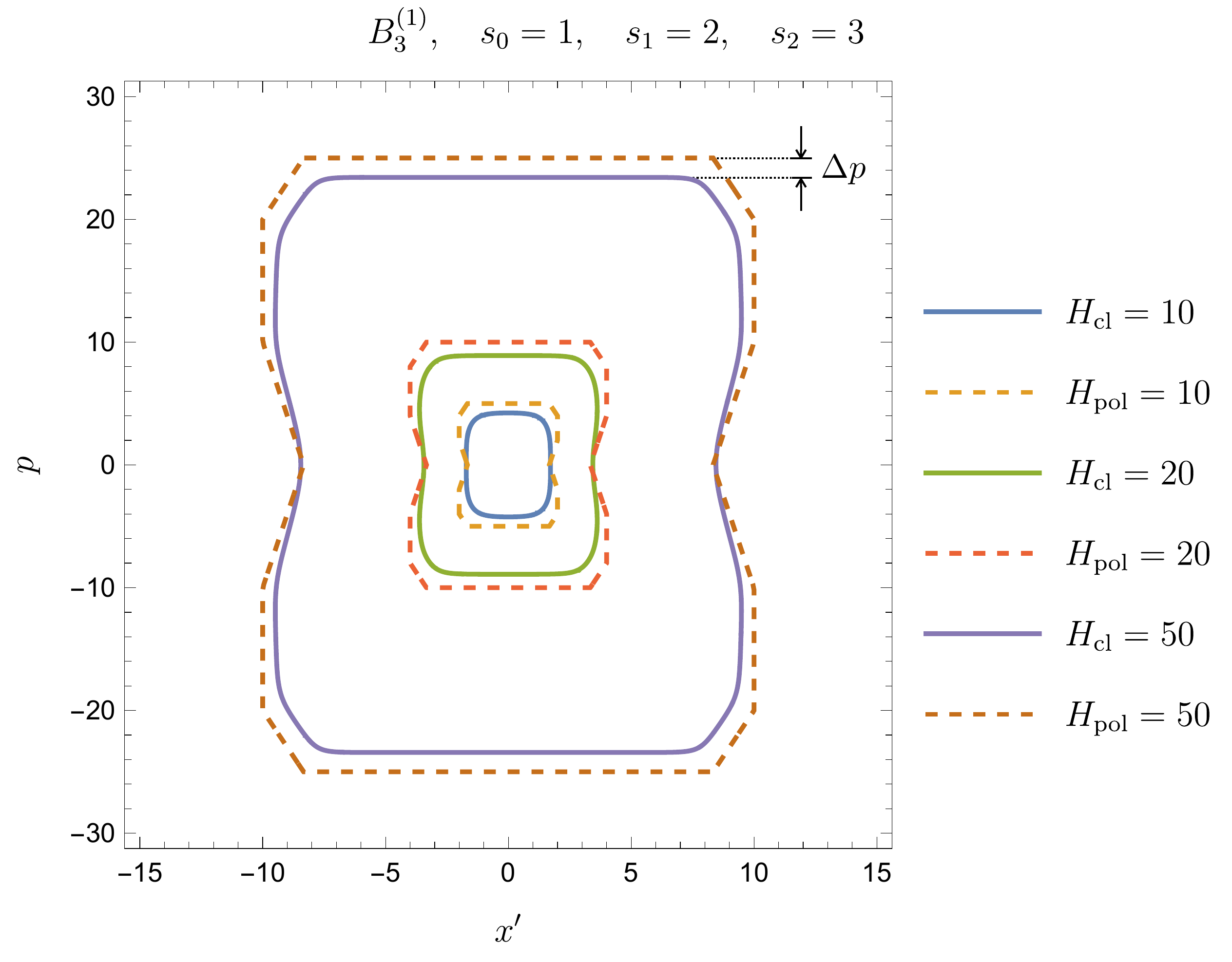}
\caption{
Comparison between the classical Fermi surface $H_{\text{cl}}(x,p)=E$ and the asymptotic polygon $H_{\text{pol}}(x,p)=E$ for the quiver $B^{(1)}_r$.
}
\label{fig_B(1)rFermisurface}
\end{center}
\end{figure}

In the classical limit $k\rightarrow 0$, the number of states can be calculated as the volume of phase space
\begin{align}
n(E)=\frac{1}{4\pi^2k}\int_{H_{\text{cl}}(x',p)\le E} dx'dp+{\cal O}(k),
\end{align}
with $H_{\text{cl}}(x',p)$ the classical Hamiltonian obtained from the density matrix \eqref{rhohatBCS} as
\begin{align}
H_{\text{cl}}(x',p)=\sum_{s=0}^{r-1}\sum_\pm\log 2\cosh\frac{p\pm\sigma_sx'}{2}-\log|2\sinh p|+\log\Bigl|\frac{p}{2}\Bigr|.
\label{Hcl}
\end{align}
In particular, when the energy $E$ is large, since $x'$ and $p$ are large at a generic point on the Fermi surface $H_{\text{cl}}(x',p)=E$, Fermi surface itself can be approximated by the polygon $H_{\text{pol}}(x',p)=E$, with $H_{\text{pol}}(x',p)$ obtained by replacing $\log 2\cosh (\cdot)$ and $\log 2\sinh (\cdot)$ in $H_{\text{cl}}(x',p)$ with $|\cdot|$ and ignoring the last term $\log|\frac{p}{2}|$
\begin{align}
H_{\text{pol}}(x',p)=\sum_{s=0}^{r-1}\sum_\pm\frac{|p\pm\sigma_sx'|}{2}-|p|.
\end{align}
See figure \ref{fig_B(1)rFermisurface} for the comparison between the classical Fermi surface and the polygon.
Therefore, the leading behavior of the classical number of states in the large $E$ limit is completely determined by the volume of the asymptotic polygon as
\begin{align}
n(E)
=\frac{1}{k}\Biggl[\frac{2E^2}{\pi^2}\biggl(\sum_{s=0}^{r-1}\frac{|\sigma_s'|-|\sigma_{s-1}'|}{\tau_s\tau_{s-1}}+\frac{|\sigma_{r-1}'|}{\tau_{r-1}^2}\biggr)+\cdots\Biggr]+{\cal O}(k).
\label{n(E)bypolygon}
\end{align}
Here we have defined $\sigma'_{-1}=0$, $\sigma'_s$ ($s=0,\cdots,r-1$) the permutation of $\sigma_s$ such that
\begin{align}
0\le \sigma'_0\le \sigma'_1\le \cdots\le \sigma'_{r-1},
\end{align}
and $\tau_s$ as
\begin{align}
\tau_s=\sum_\pm \sum_{t=0}^{r-1}|\sigma'_s\pm \sigma'_t|-2|\sigma'_s|.
\end{align}

The sub-leading correction in the small $k$ expansion can be calculated by the Wigner-Kirkwood expansion \cite{Marino:2011eh}, where the sources of the corrections to the number of states $n(E)$ are (i) the difference between the Wigner Hamiltonian $H_W(x',p)$ and $H_\text{cl}(x',p)$ and (ii) the difference between the Wigner transformation of $\theta(\hat{H}-E)$ and $\theta(H_W(x',p)-E)$.
However, since both contributions are at most ${\cal O}(E^0)$ in the large $E$ limit (see e.g.~\cite[footnote 4]{Kubo:2025dot}), we conclude that the leading behavior of $n(E)$ in the large $E$ limit is exactly given by \eqref{n(E)bypolygon}
\begin{align}
&n(E)
=2C(B^{(1)}_r)E^2
+\cdots
\end{align}
with
\begin{align}
C(B^{(1)}_r)=
\frac{1}{\pi^2 k}\biggl(\sum_{s=0}^{r-1}\frac{|\sigma_s'|-|\sigma_{s-1}'|}{\tau_s\tau_{s-1}}+\frac{|\sigma_{r-1}'|}{\tau_{r-1}^2}\biggr).
\label{CofB(h)(1)_r}
\end{align}
Plugging this into \eqref{Jfromn(E)}, we obtain the leading behavior of $J(\mu)$ in the large $\mu$ limit as \eqref{Jinlargemu}, with the coefficient $C(B^{(1)}_r)$ given by \eqref{CofB(h)(1)_r}.

Since the calculation for the other quivers $C^{(1)}_r$, $A^{(2)}_{2r-1}$, $A^{(2)}_{2r}$ and $D^{(2)}_{r+1}$ is parallel, here let us list only the essential ingredients for each model: the parametrization of the Chern--Simons levels,  the one-particle density matrix $\hat{\rho}(\Gamma)$ written in $(\hat{x}',\hat{p})$ and the coefficient $C(\Gamma)$ in the large $N$ free energy.

For $C^{(1)}_r$ quiver, by parametrizing the Chern--Simons levels as
\begin{subequations}
\begin{align}
&k_0=-2k\sigma_0,\\
&k_s=k(\sigma_{s-1}-\sigma_s),\quad (s=1,\cdots,r-1),\\
&k_r=2k\sigma_{r-1},
\end{align}
\end{subequations}
we can write the density matrix \eqref{rhohatC} as, up to a similarity transformation,
\begin{align}
\hat{\rho}(C^{(1)}_r)&=
\frac{1}{2\cosh\frac{\hat{p}+\sigma_0\hat{x}'}{2}}
\cdots
\frac{1}{2\cosh\frac{\hat{p}+\sigma_{r-1}\hat{x}'}{2}}
\frac{\hat{p}}{\pi}
\frac{1}{2\cosh\frac{\hat{p}-\sigma_{r-1}\hat{x}'}{2}}
\cdots
\frac{1}{2\cosh\frac{\hat{p}-\sigma_0\hat{x}'}{2}}
\frac{\hat{p}}{\pi},
\end{align}
with which we find the large $\mu$ expansion of the modified grand potential as \eqref{Jinlargemu}, with the coefficient $C(C^{(1)}_r)$ given by
\begin{align}
C(C^{(1)}_r)=\frac{1}{\pi^2k}\biggl(\sum_{s=0}^{r-1}\frac{|\sigma'_s|-|\sigma'_{s-1}|}{\tau_s\tau_{s-1}}+\frac{|\sigma'_{r-1}|}{\tau_{r-1}^2}\biggr),
\label{CofC(h)(1)_r}
\end{align}
with $\sigma'_{-1}=0$, $\{\sigma'_s\}$ ($s=0,\cdots,r-1$) being a permutation of $\{\sigma_s\}$ such that $|\sigma'_0|\le |\sigma'_1|\le\cdots\le|\sigma'_{r-1}|$ and
\begin{align}
\tau_s=\sum_\pm\sum_{t=0}^{r-1}|\sigma'_s\pm\sigma'_t|.
\end{align}

For $\Gamma=A^{(2)}_{2r-1}$, by parametrizing the Chern--Simons levels as
\begin{subequations}
\begin{align}
&k_0=k(-\sigma_0-\sigma_1),\quad
k_1=k(\sigma_0-\sigma_1),\\
&k_s=k(\sigma_{s-1}-\sigma_s),\quad (s=2,\cdots,r-1),\\
&k_r=2k\sigma_{r-1},
\end{align}
\end{subequations}
we can rewrite the density matrix \eqref{rhohatAodd} as
\begin{align}
\hat{\rho}(A^{(2)}_{2r-1})&=
\Bigl(
\frac{\tanh\frac{\hat{p}+\sigma_0\hat{x}'}{2}}{2}
+\frac{\tanh\frac{\hat{p}-\sigma_0\hat{x}'}{2}}{2}
\Bigr)
\frac{1}{2\cosh\frac{\hat{p}+\sigma_1\hat{x}'}{2}}
\cdots
\frac{1}{2\cosh\frac{\hat{p}+\sigma_{r-1}\hat{x}'}{2}}
\frac{\hat{p}}{\pi}\nonumber \\
&\quad
\frac{1}{2\cosh\frac{\hat{p}-\sigma_{r-1}\hat{x}'}{2}}
\cdots
\frac{1}{2\cosh\frac{\hat{p}-\sigma_1\hat{x}'}{2}}.
\end{align}
Since the classical Hamiltonian coincides with that for the quiver $\Gamma=B^{(1)}_r$ up to the ${\cal O}(\log|p|)$ terms for $|x|,|p|\gg 1$, we find the leading coefficient $C(A^{(2)}_{2r-1})$ of the large $\mu$ expansion of the modified grand potential coincides with $C(B^{(1)}_r)$ \eqref{CofB(h)(1)_r}
\begin{align}
C(A^{(2)}_{2r-1})=
\frac{1}{\pi^2 k}\biggl(\sum_{s=0}^{r-1}\frac{|\sigma_s'|-|\sigma_{s-1}'|}{\tau_s\tau_{s-1}}+\frac{|\sigma_{r-1}'|}{\tau_{r-1}^2}\biggr),
\label{CofA(h)(2)_2r-1}
\end{align}
where $\sigma'_{-1}=0$, $\sigma'_s$ ($s=0,\cdots,r-1$) the permutation of $\sigma_s$ such that $0\le |\sigma'_0|\le |\sigma'_1|\le \cdots\le |\sigma'_{r-1}|$ and $\tau_s=\sum_\pm \sum_{t=0}^{r-1}|\sigma'_s\pm \sigma'_t|-2|\sigma'_s|$.

For $\Gamma=A^{(2)}_{2r}$, we parametrize the balanced Chern--Simons levels as
\begin{subequations}
\begin{align}
&k_0=-k\sigma_0,\quad
k_s=k(\sigma_{s-1}-\sigma_s),\quad (s=1,\cdots,r-1),\\
&k_r=2k\sigma_{r-1}
\end{align}
\end{subequations}
so that the density matrix \eqref{rhohatAeven}
 becomes
\begin{align}
&\hat{\rho}(A^{(2)}_{2r})=
\frac{1}{2\cosh\frac{\hat{p}-\sigma_{r-1}\hat{x}'}{2}}
\cdots
\frac{1}{2\cosh\frac{\hat{p}-\sigma_0\hat{x}'}{2}}
\frac{1}{\hat{p}}
\frac{1}{2\cosh\frac{\hat{p}+\sigma_0\hat{x}'}{2}}
\cdots
\frac{1}{2\cosh\frac{\hat{p}+\sigma_{r-1}\hat{x}'}{2}}
\frac{2\hat{p}}{\pi}.
\end{align}
The coefficient of the large $N$ free energy is the same as that for the $\Gamma=C^{(1)}_r$ quiver matrix model \eqref{CofC(h)(1)_r}
\begin{align}
C(A^{(2)}_{2r})=\frac{1}{\pi^2k}\biggl(\sum_{s=0}^{r-1}\frac{|\sigma'_s|-|\sigma'_{s-1}|}{\tau_s\tau_{s-1}}+\frac{|\sigma'_{r-1}|}{\tau_{r-1}^2}\biggr),
\label{CofA(h)(2)_2r}
\end{align}
with $\sigma'_{-1}=0$, $\{\sigma'_s\}$ ($s=0,\cdots,r-1$) being a permutation of $\{\sigma_s\}$ such that $|\sigma'_0|\le |\sigma'_1|\le\cdots\le|\sigma'_{r-1}|$ and $\tau_s=\sum_\pm\sum_{t=0}^{r-1}|\sigma'_s\pm\sigma'_t|$.

For $\Gamma=D^{(2)}_{r+1}$, we parametrize the Chern--Simons levels as
\begin{subequations}
\begin{align}
&k_0=-k\sigma_0,\quad
k_s=k(\sigma_{s-1}-\sigma_s),\quad (s=1,\cdots,r-1),\\
&k_r=k\sigma_{r-1}
\end{align}
\end{subequations}
so that the density matrix \eqref{rhohatD} becomes
\begin{align}
&\hat{\rho}=
\frac{4}{\hat{p}}
\frac{1}{2\cosh\frac{\hat{p}+\sigma_{0}\hat{x}'}{2}}
\cdots
\frac{1}{2\cosh\frac{\hat{p}+\sigma_{r-1}\hat{x}'}{2}}
\frac{1}{\hat{p}}
\frac{1}{2\cosh\frac{\hat{p}-\sigma_{r-1}\hat{x}'}{2}}
\cdots
\frac{1}{2\cosh\frac{\hat{p}-\sigma_0\hat{x}'}{2}}.
\end{align}
The coefficient of the large $N$ free energy is the same as that for the $\Gamma=C^{(1)}_r$ quiver matrix model \eqref{CofC(h)(1)_r}
\begin{align}
C(D^{(2)}_{r+1})=\frac{1}{\pi^2k}\biggl(\sum_{s=0}^{r-1}\frac{|\sigma'_s|-|\sigma'_{s-1}|}{\tau_s\tau_{s-1}}+\frac{|\sigma'_{r-1}|}{\tau_{r-1}^2}\biggr),
\label{CofD(h)(2)_r+1}
\end{align}
with $\sigma'_{-1}=0$, $\{\sigma'_s\}$ ($s=0,\cdots,r-1$) being a permutation of $\{\sigma_s\}$ such that $|\sigma'_0|\le |\sigma'_1|\le\cdots\le|\sigma'_{r-1}|$ and $\tau_s=\sum_\pm\sum_{t=0}^{r-1}|\sigma'_s\pm\sigma'_t|$.

\subsection{Numerical check}

In table \ref{table_FermigasvslargeNsaddle} we also compare the coefficient of $N^{3/2}$ in the large $N$ free energy \eqref{largeNfreeenergyfromFG} with the same coefficient determined by the large $N$ saddle point approximation \cite{Herzog:2010hf}.\footnote{
When we complexify the integration variables in the large $N$ saddle point approximation, we have treated the contributions to the effective action \eqref{Seff} from the Vandermonde factors with $\beta=1$ as
\begin{align}
-\sum_{i<j}^{a_sN}\log\biggl|2\sinh\frac{x^{(s)}_i-x^{(s)}_j}{2}\biggr|
\rightarrow
-\sum_{i<j}^{a_sN}\log\biggl(\operatorname{sgn}(\operatorname{Re}[x^{(s)}_i-x^{(s)}_j])2\sinh\frac{x^{(s)}_i-x^{(s)}_j}{2}\biggr).
\end{align}
}
For simplicity we consider only a few examples where the Chern--Simons levels satisfy $k_s=-k_{s'}$ for all pairs $(s,s')$ of nodes related under the reflection about the symmetry axis of the quiver so that the effective action $S_\text{eff}(\Gamma)$ \eqref{Seff} of the quiver matrix model is manifestly real when the complexified configuration of the integration variables satisfy $x^{(s)}_i=(x^{(s')}_i)^*$.
We skip $A^{(2)}_{2r}$, where there are no such symmetric choice of the Chern--Simons levels except $k_0=\cdots=k_r=0$.
We find the coefficient of $N^{3/2}$ from the large $N$ saddle point approximation for finite valuess of $k$ shows good agreement with the analytic expression we have obtained from the Fermi gas formalism, \eqref{largeNfreeenergyfromFG} with \eqref{CofB(h)(1)_r}, \eqref{CofC(h)(1)_r}, \eqref{CofA(h)(2)_2r-1} and \eqref{CofD(h)(2)_r+1}.
This also supports the validity of our Fermi gas results for finite $k$.
\begin{table}
\begin{center}
\begin{align*}
\begin{array}{|c|c|c|c|c|}
\hline
\multirow{2}{*}{$\Gamma$}&\multirow{2}{*}{$(k;\sigma_0,\cdots,\sigma_{r-1})$}&\multirow{2}{*}{Fermi gas}&\multicolumn{2}{c|}{\text{large }N\text{ saddle}}\\ \cline{4-5}
                         &                                                   &                          &\text{numerical}&\text{analytic}\\ \hline\hline
B^{(1)}_3&(1;1,0,0)             &\frac{8 \pi}{3 \sqrt{3}}=4.837        &4.487\,\,(N=30,40,50,60)                &\frac{8\pi}{3\sqrt{3}}\\ \hline
B^{(1)}_4&(1;1,0,0,0)           &2\pi=6.283                            &5.896\,\,(N=77,78,79,80)                &2\pi\\ \hline
B^{(1)}_r&(k;1,0,\cdots,0)      &\frac{4\pi(r-1)\sqrt{k}}{3\sqrt{r}}   &-                                       &\frac{4\pi(r-1)\sqrt{k}}{3\sqrt{r}}\\ \hline\hline
C^{(1)}_2&(\frac{1}{2};-1,-1)   &\frac{4 \pi}{3}=4.189                 &3.939\,\,(N=130,140,150,160)            &\frac{4 \pi}{3}\\ \hline 
C^{(1)}_3&(1;-1,-2,-1)          &\frac{16 \pi}{\sqrt{23}}=10.48        &10.04\,\,(N=60,70,80,100)               &\text{-}\\ \hline
C^{(1)}_r&(k;-1,\cdots,-1)      &\frac{2\pi r\sqrt{2k}}{3}             &-                                       &\frac{2\pi r\sqrt{2k}}{3}\\ \hline\hline
A^{(2)}_5&(1;1,0,0)             &\frac{8 \pi}{3 \sqrt{3}}=4.837        &4.494\,\,(N=73,74,75,76)                &\text{-}\\ \hline \hline
D^{(2)}_3&(1;-1,-1)             &\frac{2 \pi}{3}=2.094               &2.018\,\,(N=140,160,180,200)            &\text{-}\\ \hline
D^{(2)}_4&(1;-2,-3,-2)          &\frac{7\pi\sqrt{3}}{\sqrt{62}}=4.837&4.741\,\,(N=214,216,218,220)            &\text{-}\\ \hline
\end{array}
\end{align*}
\caption{
Comparison of the coefficient of $N^{3/2}$ of the free energy $-\log Z(\Gamma;N)$ in the large $N$ limit.
The values in the column ``Fermi gas'' are given by $\frac{2}{3\sqrt{C(\Gamma)}}$.
The values in the column ``large $N$ saddle'' are obtained by writing the integrand of the quiver matrix model as $e^{-S_\text{eff}(\Gamma)}$ \eqref{Seff} and approximating the free energy by $S_\text{eff}(\Gamma)$ evaluated at the saddle point determined by $\frac{\partial S_\text{eff}(\Gamma)}{\partial x^{(s)}_i}=0$ for all $s\in\Gamma_0$ and $i=1,\cdots,N_s$.
For the subcolumn ``numerical'', we have solved the saddle point equations numerically by FindRoot method in Mathematica with the option WorkingPrecision$\rightarrow$200, and determined the coefficient of $N^{3/2}$ by fitting the values of $S_\text{eff}(\Gamma)$ at the values of $N$'s listed in the table by $c_1N^{3/2}+c_2N+c_3N^{1/2}$ with the fitting parameters $(c_1,c_2,c_3)$.
For the subcolumn ``analytic'', we have solved the saddle point equations analytically in the large $N$ limit with the ansatz \eqref{HKPTansatz}.
}
\label{table_FermigasvslargeNsaddle}
\end{center}
\end{table}

\subsection{Comment on sub-leading corrections}
In the Fermi gas formalism of of the circular quiver Chern--Simons theory on $A^{(1)}_r$ or $D^{(1)}_r$ quiver with balanced levels, the large $E$ expansion of the number of states $n(E)$ consists only of the quadratic term and the constant term except the non-perturbatively small correcitons
\begin{align}
n(E)=CE^2+B-\frac{\pi^2 C}{3}+{\cal O}(e^{-\# E}),
\end{align}
from which it follows that the matrix model coincides with the Airy function up to the non-perturbative corrections in $1/N$ \cite{Marino:2011eh,Moriyama:2015jsa}.

This behavior of $n(E)$ comes from the fact that in these models the Fermi surface deviate from the asymptotic polygon only around the vertices of the polygon.
For $\Gamma=B^{(1)}_r$, on the other hand, as we have shown in figure \ref{fig_B(1)rFermisurface}, the deviation is extended in one-dimension.
This is due to the $\log|p|$ term in the classical Hamiltonian \eqref{Hcl}, and hence also happens for $\Gamma=C^{(1)}_r$, $A^{(2)}_{2r-1}$ and $D^{(2)}_{r+1}$.
The size of the deviation depends on $E$ as $\Delta p\sim \log E$.
As a result, the number of state receives a sub-leading correction of the order $E\log E$.
Hence we conclude that the large $N$ expansion of the matrix model does not coincides with the Airy function for $\Gamma=B^{(1)}_r$, $C^{(1)}_r$, $A^{(2)}_{2r-1}$ and $D^{(2)}_{r+1}$.
On the other hand, for $\Gamma=A^{(2)}_{2r}$ the factor $\hat{p}$ and $\frac{1}{\hat{p}}$ in the density matrix cancel and hence the Fermi surface is indistinguishable from that of an affine $A$ type quiver in the classical limit.
Hence the current analysis is insufficient to conclude whether the Airy structure breaks down or not, which we leave as a future work.

\section{Derivation of quiver matrix models}
\label{sec_derivationQMM}

Let us discuss a physical origin of the quiver Chern--Simons matrix model discussed above.
As discussed above, we consider the $\beta$-deformation to realize quiver matrix models.
It was pointed out by Brini, Marino, and Stevan~\cite{Brini:2010fc} that the $\beta$-deformation of the Chern--Simons matrix model is obtained via the $\beta$-deformation of the Stieltjes--Wigert matrix model \cite{Marino:2002fk,Tierz:2002jj}.
Another deformation is the so-called refined Chern--Simons theory given by Aganagic and Shakirov via the geometric transition from the refined topological string~\cite{Aganagic:2011sg}, which corresponds to the Macdonald deformation rather than the Jack deformation.
Such a Macdonald deformation is indeed obtained from the localization computation of three-dimensional gauge theroy on $\mathbb{D}^2 \times \mathbb{S}^1$~\cite{Yoshida:2014ssa}.
The vector multiplet contribution of $\mathcal{N}=4$ $\mathrm{U}(N)$ theory is given by
\begin{align}
    z_\text{vec} = \prod_{1 \le i \neq j \le N} \frac{(z_i/z_j;q)_\infty}{(tz_i/z_j;q)_\infty} = \prod_{1 \le i \neq j \le N} \exp \left( - \sum_{n=1}^\infty \frac{1-t^n}{1-q^n} \frac{(z_i/z_j)^n}{n} \right) \, ,
\end{align}
where $|q|<1$ is the rotation fugacity and $t$ is the (exponentiated) adjoint mass.
Tuning the parameter $t = q^{\beta/2}$ and taking the limit $q \to 1$,
\footnote{The limit $q \to 1$ is indeed the Nekrasov--Shatashvili (NS) limit~\cite{Nekrasov:2009rc}. In the current case, however, we also scale the adjoint mass, while the standard NS limit keeps the adjoint mass finite.}
we have
\begin{align}
    z_\text{vec}\Big|_{t = q^{\beta/2}} & = \prod_{1 \le i \neq j \le N} \exp \left( - \sum_{n=1}^\infty \frac{1-q^{n\beta/2}}{1-q^n} \frac{(z_i/z_j)^n}{n} \right) \nonumber \\
    & \xrightarrow{q \to 1} \prod_{1 \le i \neq j \le N} \exp \left( - \frac{\beta}{2} \sum_{n=1}^\infty \frac{(z_i/z_j)^n}{n} \right) = \prod_{1 \le i \neq j \le N} \left( 1 - \frac{z_j}{z_i} \right)^{\beta/2} \, .
\end{align}
Since $z = (z_1,\ldots,z_N)$ takes a value in the Cartan torus, we write $z_j = \exp(i x_j)$.
Then, the vector multiplet contribution is reduced to the $\beta$-deformed Vandermonde determinant,
\begin{align}
    z_\text{vec} \xrightarrow[q \to 1]{t = q^{\beta/2}} \prod_{1 \le i < j \le N} \left( 2 i \sin (x_i - x_j) 2 i \sin (x_j - x_i) \right)^{\beta/2} = \prod_{1 \le i < j \le N} \left| 2 \sin (x_i - x_j) \right|^\beta \, .
\end{align}
This is the trigonometric version of the $\beta$-deformation, and the integral is taken over the Cartan torus, which is compact.
In order to obtain the hyperbolic version, we may use the following formula for $g = \log q$~\cite{Okuda:2004mb,Aganagic:2011sg},
\begin{align}
    q^{m^2/2} = \oint \left( \sum_{n \in \mathbb{Z}} q^{n^2/2} z^n \right) \frac{dz}{2\pi i z^{m+1}} = \int_{-\infty}^\infty e^{-\frac{1}{2g} x^2} e^{mx} \frac{dx}{\sqrt{2\pi g}} \, ,
\end{align}
where the integrand of the first integral is the Jacobi theta function $\vartheta_3(z;q)$.
Then, we have the hyperbolic version of the $\beta$-deformation with the Gaussian weight.

Moreover, in order to obtain the quiver generalization of the Chern--Simons matrix model, we incorporate additional adjoint fields accordingly.
For example, we have
\begin{align}
    \prod_{1 \le i \neq j \le N} \frac{(z_i/z_j;q)_\infty}{(tz_i/z_j;q)_\infty} \frac{(tz_i/z_j;q)_\infty}{(t^2z_i/z_j;q)_\infty} = \prod_{1 \le i \neq j \le N} \frac{(z_i/z_j;q)_\infty}{(t^2z_i/z_j;q)_\infty} \xrightarrow[q \to 1]{t = q^{\beta/2}} \prod_{1 \le i \neq j \le N} \left( 1 - \frac{z_j}{z_i} \right)^{\beta} \, .
\end{align}
In this way, we can effectively assign node-dependent adjoint masses.
This construction is interpreted as a reduction of the quiver W-algebra formalism~\cite{Kimura:2015rgi,Kimura:2016dys,Kimura:2017hez}, which realizes non-simply-laced quiver gauge theory by imposing node-dependent $\Omega$-background parameters since one of them is identified with the adjoint mass in the three-dimensional setup.

\section{Discussion}
\label{sec_discussion}

In this paper we have considered matrix models associated with affine $BC$ and twisted affine $AD$ type quivers, which generalize the affine $A$ (i.e.~circular) and $D$ matrix models studied intensively in the previous works.
As explained in section \ref{sec:aff_quiver_matrix_model}, these quiver matrix models (including those for affine $AD$) are defined by assigning the Vandermonde determinant factor to each node and the bifundamental factor to each edge, whose powers are given by the components of the symmetrized Cartan matrix associated with the quiver diagram.
If we further choose the ranks $N_s$ of the nodes proportional to the Kac labels $a_s$ (which we call balanced) and the integration measures to be the Fresnel factors whose Chern--Simons levels $k_s$ satisfy the level balance condition \eqref{level_balance}, the large $N$ saddle point approximation shows that the free energy of the matrix model scales as $N^{3/2}$ in the large $N$ limit, which is characteristic of the M2-brane worldvolume theory.

In particular, we have extended the Fermi gas formalism, which has been established for the affine $AD$ type quivers, to these new quiver matrix models with balanced ranks.
Namely, we have shown that the grand canonical sum of the matrix model $Z(\Gamma;N)$ for each of these quivers $\Gamma$ with respect to the rank $N$ is given by the Fredholm Pfaffian of an operator $\hat{\rho}(\Gamma)$ of one-dimensional quantum mechanics.
This structure of Fredholm Pfaffian is a natural extension of the Fredholm determinant for the grand canonical sum of the affine $AD$ type quiver matrix models. 

As an application, we have determined the leading coefficient of $N^{3/2}$ in the large $N$ free energy for the Fresnel measures with general Chern--Simons levels $\{k_s\}$ under the level balance condition \eqref{level_balance}.
We have also confirmed that the coefficient agrees with the numerical result of the large $N$ saddle point analysis for several examples.
In the Fermi gas formalism, this coefficient of $N^{3/2}$ in the free energy is related to the growth rate of the Fermi see volume determined by the Fermi surface $-\log \hat{\rho}=E$ in the phase space.
Although this picture relies on the semiclassical expansion with respect to $\hbar\sim k_s$, and in particular we have focused on the leading contribution, the agreement with the large $N$ saddle point approximation supports the validity of our result also for finite $k_s$.

There are several related problems which we hope to address in the future.
First, although in this paper we have determined only the coefficients of the leading $N^{3/2}$ term in the free energy, it is possible to extend our analysis to the sub-leading corrections in the $1/N$ expansion by using the Fermi gas formalism.
In particular, our analysis suggests at least for $B^{(1)}_r$, $C^{(1)}_r$, $A^{(2)}_{2r-1}$ and $D^{(2)}_{r+1}$ quiver that the $1/N$ corrections are not given by the Airy function, unlike in the affine $AD$ type quivers.
We hope to reveal the structure of all order $1/N$ corrections in these new quivers which alternates the Airy function.

Our construction of the affine $BC$ and twisted affine $AD$ quivers is based on the $\beta$-deformation of the boundary nodes.
A related construction has been discussed in \cite{Gulotta:2011vp}, which assigns O/USp gauge groups to the boundary nodes instead of the $\beta$-deformation. 
It would be interesting to construct the Fermi gas formalism also for these O/USp constructions of the quiver matrix models, where one would be able to compare various outcomes of the Fermi gas formalism with the physical insights from such as the field theory dualities and holography.
It would be also interesting to combine the $\beta$-deformation and the O/USp construction.

While we have focused on the quiver matrix models whose ranks strictly obey the balance condition $N_s=a_sN$, the long-range force cancellation is still valid even if the rank of each node is shifted by an ${\cal O}(1)$ factor.
Moreover, the Fermi gas formalism for affine $AD$ type quivers persists at least for some special types of the rank deformations \cite{Kubo:2020qed,Kubo:2024raz}.
More concretely, the Fermi gas formalism for such rank deformed matrix models can be constructed in two different ways: (i) the closed string formalism \cite{Awata:2012jb}, where the grand canonical sum is given by the Fredholm determinant/Pfaffian whose density matrix $\hat{\rho}$ is modified by the relative ranks, and (ii) the open string formalism \cite{Matsumoto:2013nya}, where the Fredholm determinant/Pfaffian is multiplied by an extra factor of finite size determinant/Pfaffian, which is similar to what we have found for the $D^{(2)}_{r+1}$ quivers with odd ranks \eqref{XioddD(2)r+1final}.
It would be interesting also to generalize our construction of the Fermi gas formalism for affine $BC$ and twisted affine $AD$ type quivers to those with rank deformations.
It would also be interesting to investigate whether there are restrictions on the relative ranks, or the duality ($\infty$-ality) among the matrix models with different relative ranks \cite{Assel:2014awa,Honda:2020uou,Furukawa:2020cjp,Furukawa:2021pll,Furukawa:2022uub,Moriyama:2023pxd,Moriyama:2024bbn,Moriyama:2026eat}, which in physical setups would correspond to the Seiberg-like dualities.

The supersymmetric gauge theory on affine $D$ type quiver with unitary gauge groups is sometimes mirror to a circular quiver theory with $\text{USp}$ gauge groups and hence the $S^3$ partition functions of the two theories coincide with each other.
For example, the ${\cal N}=4$ supersymmetric gauge theory on $D^{(1)}_{\ell+3}$ quiver with balanced ranks of the unitary gauge groups and one fundamental hypermultiplet on one of the four affine nodes is mirror to the $\text{USp}(2N)$ gauge theory with an antisymmetric hypermultiplet and $2\ell$ fundamental half hypermultiplets \cite{Kapustin:1998fa,Hanany:1999sj}.
In \cite{Mezei:2013gqa} the Fermi gas formalism of the latter theory was constructed for the cases without mass deformations, where grand canonical sum is given by a Fredholm determinant instead of the Fredholm Pfaffian, whose equivalence with the Fredholm Pfaffian formulation was also shown in \cite{Assel:2015hsa}.
It would be interesting to investigate the similar alternative Fermi gas formalisms also in our setups and argue their physical interpretation.

In \cite{Grassi:2014uua,Bonelli:2017gdk} the grand canonical sum of the $S^3$ partition function of the $\text{U}(N)_k\times \text{U}(N+M)_{-k}$ ABJ theory was found to satisfy bilinear difference relation with respect to the relative rank $M$.
The same structure was also discovered for several other affine $A$ type quivers \cite{Nosaka:2020tyv,Bonelli:2022dse,Moriyama:2023mjx,Moriyama:2023pxd,He:2025zxk}, some of which can also be written without rank deformations by using the Seiberg-like duality.
It would be interesting to investigate similar structures also for the affine $D$ type quivers as well as the affine $BC$ and twisted affine $AD$ type quivers with or without rank deformations.

\acknowledgments

This work of TK was supported by EIPHI Graduate School (No.~ANR-17-EURE-0002) and the Bourgogne-Franche-Comté region.
The work of TN was supported by the Startup Funding no.~2302-SRFP-2024-0012 of Shanghai Institute for Mathematics and Interdisciplinary Sciences.

\appendix

\section{Notation of one-dimensional quantum mechanics}
\label{app_1dQM}

For the one-dimensional quantum mechanical system with $[\hat{x},\hat{p}]=i\hbar$ with $\hbar=2\pi$, we fix the normalization of the position eigenstates $|x\rangle$ and the momentum eigenstates $|p\rangle\!\rangle$ as
\begin{subequations}
\label{1dqmnotation}
\begin{align}
\hat{x}|x\rangle&=x|x\rangle,\quad \hat{p}|p\rangle\!\rangle=p|p\rangle\!\rangle,\\
\langle x|x'\rangle&=2\pi\delta(x-x'),\quad \int\frac{dx}{2\pi}|x\rangle\langle x|=1,\\
\langle\!\langle p|p'\rangle\!\rangle&=2\pi\delta(p-p'),\quad \int\frac{dp}{2\pi}|p\rangle\!\rangle\langle\!\langle p|=1,\\
\langle x|p\rangle\!\rangle&=e^{\frac{ixp}{2\pi}},\quad
\langle\!\langle p|x\rangle=e^{-\frac{ixp}{2\pi}}.
\end{align}
\end{subequations}

Under these normalizations we have the following formulas for the position matrix elements of the momentum operators
\begin{subequations}
\label{sgnFourier}
\begin{align}
\langle x|\frac{1}{2\cosh\frac{\hat{p}}{2}}|y\rangle&=\frac{1}{2\cosh\frac{x-y}{2}},\\
\langle x|\Bigl(-\frac{i\tanh\frac{\hat{p}}{2}}{2}\Bigr)|y\rangle&=\frac{1}{2\sinh\frac{x-y}{2}},\\
\langle x|\Bigl(-\frac{i\hat{p}}{4\pi\cosh\frac{\hat{p}}{2}}\Bigr)|y\rangle&=\frac{\sinh\frac{x-y}{2}}{(2\cosh\frac{x-y}{2})^2},\\
\langle x|\Bigl(-\frac{2i}{\hat{p}}\Bigr)|y\rangle&=\text{sgn}(x-y),
\end{align}
\end{subequations}
which we use in section \ref{sec_Fremigas}.

We also define the transposition of a quantum mechanical operator $\hat{\cal O}$ by $\langle x|\hat{\cal O}|y\rangle=\langle y|\hat{\cal O}^{\rmT}|x\rangle$.
More explicitly, we have
\begin{align}
(
f_1(\hat{x})
g_1(\hat{p})
\cdots
f_n(\hat{x})
g_n(\hat{p})
)^{\rmT}
=
g_n(-\hat{p})
f_n(\hat{x})
\cdots
g_1(-\hat{p})
f_1(\hat{x}).
\label{Ot}
\end{align}

\section{Formulas}
\label{app_formulas}

In this appendix we list basic formulas we have used in section \ref{sec_Fremigas} in the derivation of the Fermi gas formalism for the affine quiver matrix models.

\subsection{Symmetric products as determinant/Pfaffian}
For the products of hyperbolic functions, we have the Cauchy determinant formula
\begin{align}
&\frac{\prod_{i<j}^N2\sinh\frac{x_i-x_j}{2}
\prod_{i<j}^N2\sinh\frac{y_i-y_j}{2}
}{
\prod_{i,j}^N2\cosh\frac{x_i-y_j}{2}
}
=\det\Bigl(\frac{1}{2\cosh\frac{x_i-y_j}{2}}
\Bigr)=\det\Bigl(\langle x_i|\frac{1}{2\cosh\frac{\hat{p}}{2}}|y_j\rangle\Bigr),\label{Cauchydetsinh} \\
&\frac{\prod_{i<j}^N2\sinh\frac{x_i-x_j}{2}
\prod_{i<j}^N2\sinh\frac{y_i-y_j}{2}
}{
\prod_{i,j}^N2\sinh\frac{x_i-y_j}{2}
}
=(-1)^{\frac{N(N-1)}{2}}\det\Bigl(\frac{1}{2\sinh\frac{x_i-y_j}{2}}
\Bigr)\nonumber \\
&\quad\quad\quad\quad\quad\quad\quad\quad\quad\quad\quad\quad\quad\quad\quad
=(-1)^{\frac{N(N-1)}{2}}\det\Bigl(\langle x_i|\Bigl(-\frac{i\tanh\frac{\hat{p}}{2}}{2}\Bigr)|y_j\rangle\Bigr).\label{Cauchydetsinhsinh}
\end{align}
Here we have also rewritten the matrix element in the notation of one-dimensional quantum mechanics introduced in appendix \ref{app_1dQM}.
The second formula is obtained by shifting $y_i$ by $\pi i$ in the first formula.

For the product of $\text{sgn}(x_i-x_j)$, we have the following formula
\begin{align}
\prod_{i<j}^{2N}\text{sgn}(x_i-x_j)=
\operatorname*{pf}_{2N\times 2N}[\text{sgn}(x_i-x_j)]=\operatorname*{pf}_{2N\times 2N}\Bigl(\langle x_i|\Bigl(-\frac{2i}{\hat{p}}\Bigr)|x_j\rangle\Bigr)
\label{VdMlikeforsgn}
\end{align}
when the number of variables is even.
Here $\operatorname{pf}(A)$ is Pfaffian which is defined for an $2n\times 2n$ anti-symmetric matrix $A$ as
\begin{align}
\operatorname{pf}(A)=(-1)^{\frac{n(n-1)}{2}}\frac{1}{2^nn!}\sum_{\sigma\in S_{2n}}(-1)^\sigma \prod_{i=1}^nA_{\sigma(i),\sigma(n+i)}.
\end{align}
When the number of variables is odd, we have
\begin{align}
\prod_{i<j}^{2N-1}\text{sgn}(x_i-x_j)&=
\operatorname*{pf}_{2N \times 2N}
\begin{pmatrix}
[\text{sgn}(x_i-x_j)]_{i,j}^{(2N-1)\times (2N-1)}&\begin{bmatrix}
1\\
\vdots\\
1\\
\end{bmatrix}^{(2N-1)\times 1}\\
[-1 \cdots -1]^{1\times (2N-1)}&0
\end{pmatrix}\nonumber \\
&
=
\operatorname{pf}
\begin{pmatrix}
\Bigl[\langle x_i|\Bigl(-\frac{2i}{\hat{p}}\Bigr)|x_j\rangle\Bigr]_{i,j}^{(2N-1)\times (2N-1)}&
[\langle x_i|0\rangle\!\rangle]_{i=1}^{2N-1}\vspace{0.2cm} \\
[-\langle\!\langle 0|x_j\rangle]_{j=1}^{2N-1}&0
\end{pmatrix}.
\label{VdMlikeforsgnNodd}
\end{align}

\subsection{Convolution of determinant/Pfaffian}

For the convolution across two determinants, we have the Andréief identity
\begin{align}
\frac{1}{N!}\int d^Nx\det(f_i(x_j))\det(g_j(x_i))=\det\biggl(\int dxf_i(x)g_j(x)\biggr).
\label{Andreief}
\end{align}

For the self-convolution mediated by a Pfaffian, we have a formula known as De Bruijn's formula (see e.g.~\cite{Babinet:2022nxg})
\begin{align}
\frac{1}{(2N)!}\int d^{2N}x\det_{2N\times 2N}[f_i(x_j)]\operatorname*{pf}_{2N\times 2N}A(x_i,x_j)
=\operatorname*{pf}_{2N\times 2N}\biggl[\int f_i(x)A(x,y)f_j(y)dxdy\biggr].
\label{Bruijn}
\end{align}
This formula can also be generalized by adding extra blocks (see e.g.~\cite{Babinet:2022nxg}):
\begin{align}
&\frac{1}{N!M!}\int d^{N}x d^My \det_{(N+M)\times (N+M)}
\begin{pmatrix}
[f_a(x_j)]_{a,j}^{(N+M)\times N}&[g_b(y_s)]_{a,s}^{(N+M)\times M}\end{pmatrix}\nonumber \\
&\quad\times
\operatorname*{pf}_{(N+M)\times (N+M)}
\begin{pmatrix}
[A(x_i,x_j)]_{i,j}^{N\times N}&[S(x_i,y_s)]_{i,s}^{N\times M}\vspace{0.2cm} \\
-[S(x_j,y_r)]_{r,j}^{M\times N}&[B(y_r,y_s)]_{r,s}^{M\times M}
\end{pmatrix}\nonumber \\
&
=\operatorname*{pf}_{(N+M)\times (N+M)}\Bigl(
\int dxdx'f_a(x)A(x,x')f_b(x')
+\int dxdyf_a(x)S(x,y)g_b(y)\nonumber \\
&\quad\quad\quad\quad\quad\quad\quad\quad\quad\quad -\int dydxg_a(y)S(x,y)f_b(x)+\int dydy'g_a(y)B(y,y')g_b(y')
\Bigr),
\end{align}
where $N+M\in 2\mathbb{N}$ ($N$ and $M$ are not necessary to be even).
In particular, by choosing $M=1$, 
$f_{N+1}(x)=0$, $g_a(y)=\delta_{a,N+1}$, $S(x,y)=1$, $B(y,y')=0$ and the integration domain of $y$ to be of length $1$, we obtain
\begin{align}
&\frac{1}{N!}\int d^Nx\det_{N\times N}[f_a(x_j)]
\operatorname*{pf}_{(N+1)\times (N+1)}
\begin{pmatrix}
[A(x_i,x_j)]_{i,j}^{N\times N}&
\begin{bmatrix}
1\\
\vdots\\
1
\end{bmatrix}^{N\times 1}\\
[-1 \cdots -1]^{1\times N}
&0
\end{pmatrix}\nonumber \\
&
=\operatorname*{pf}_{(N+1)\times (N+1)}
\begin{pmatrix}
[
\int dx dx'f_a(x)A(x,x')f_b(x')]_{a,b}^{N\times N}& [\int dx f_a(x)]_{a=1}^N\vspace{0.2cm} \\
[-\int dx f_b(x)]_{b=1}^N&0
\end{pmatrix},
\label{BruijnN+1xN+1}
\end{align}
which we use in appendix \ref{app_sec_D(h)(2)r+1}.

For $M=N$, $A(x,x')=B(y,y')=0$ and $S(x,y)=\delta(x-y)$, we obtain the following formula which is a Pfaffian analogue of the Andréief identity \eqref{Andreief}
\begin{align}
&\frac{1}{N!}
\int
d^Nx
\det\Bigl(
\left[f_a(x_j)\right]_{a,j}^{2N\times N}\,\,
\left[g_{a}(x_j)\right]_{a,j}^{2N\times N}
\Bigr)\nonumber \\
&
=(-1)^{\frac{N(N-1)}{2}}\operatorname{pf}\Bigl(
\int dx(f_a(x)g_b(x)-g_a(x)f_b(x))
\Bigr).
\label{CBPfaffian}
\end{align}

For the generating function of the self-convolution of $N$-variable Pfaffians with respect to $N$, we have the following formula (see e.g.~\cite{Kubo:2024raz})
\begin{align}
&\sum_{N=0}^\infty u^N \frac{(-1)^{\frac{N(N-1)}{2}}}{N!} \int d^Nx\operatorname{pf}
\begin{pmatrix}
[A(x_i,x_j)]_{i,j}^{N\times N}
&[B(x_i,x_j)]_{i,j}^{N\times N}\vspace{0.2cm} \\
[-B(x_j,x_i)]_{i,j}^{N\times N}
&[C(x_i,x_j)]_{i,j}^{N\times N}
\end{pmatrix}\nonumber \\
&
=\sqrt{\operatorname{Det}\begin{pmatrix}uA&1+uB\\
-1-uB^{\rmT}&uC
\end{pmatrix}},
\label{FredholmPfaffian}
\end{align}
which is analogous to the Fredholm determinant formula $\sum_{N=0}^\infty \frac{u^N}{N!}\int d^Nx\det A(x_i,x_j)=\operatorname{Det}(1+uA)$ (see e.g.~\cite{Marino:2011eh}).
This formula can be generalized by adding extra blocks as (see e.g.~\cite{Kubo:2024raz})
\begin{align}
&\sum_{N=0}^\infty u^N \frac{(-1)^{\frac{N(N-1)}{2}}}{N!} \int d^Nx \operatorname{pf}
\begin{pmatrix}
[A(x_i,x_j)]_{i,j}^{N\times N}&[B(x_i,x_j)]_{i,j}^{N\times N}&[v_s(x_i)]_{i,s}^{N\times 2L}\vspace{0.2cm} \\
[-B(x_j,x_i)]_{i,j}^{N\times N}&[C(x_i,x_j)]_{i,j}^{N\times N}&[w_s(x_i)]_{i,s}^{N\times 2L}\vspace{0.2cm} \\
[-v_r(x_j)]_{r,j}^{2L\times N}&[-w_r(x_j)]_{r,j}^{2L\times N}&[\alpha_{r,s}]_{r,s}^{2L\times 2L}
\end{pmatrix}\nonumber \\
&
=(-1)^{\frac{L(L-1)}{2}}\sqrt{\operatorname{Det}\begin{pmatrix}uA&1+uB&[uv_s]_{s=1}^{2L}\\
-1-uB^{\rmT}&uC&[uw_s]_{s=1}^{2L}\\
[-v_r]_{r=1}^{2L}&[-w_r]_{r=1}^{2L}&[\alpha_{r,s}]_{r,s}^{2L\times 2L}
\end{pmatrix}
},
\label{FredholmPfaffianrankdeform}
\end{align}
which we use in appendix \ref{app_sec_D(h)(2)r+1}.

\section{$D^{(2)}_{r+1}$ quiver matrix model with odd ranks}
\label{app_sec_D(h)(2)r+1}

In this appendix we construct the Fermi gas formalism of $D^{(2)}_{r+1}$ type quiver matrix model with balanced ranks $Z(D^{(2)}_{r+1};N)$ \eqref{sec_Fremigas} for odd $N$, which we have skipped in section \ref{sec:aff_quiver_matrix_model}.
First let us define the building block ${\cal O}_{\circ\!\Leftarrow,\text{odd}}^{0,1}[x^{(1)}_i]$ analogous to ${\cal O}_{\circ\!\Leftarrow}^{0,1}[x^{(1)}_i]$
\begin{align}
{\cal O}_{\circ\!\Leftarrow,\text{odd}}^{0,1}[x^{(1)}_i]=\frac{\prod_{i<j}^{2N-1}|2\sinh\frac{x^{(0)}_i-x^{(0)}_j}{2}|\prod_{i<j}^{2N-1}2\sinh\frac{x^{(1)}_i-x^{(1)}_j}{2}}{\prod_{i,j=1}^{2N-1}2\cosh\frac{x^{(0)}_i-x^{(1)}_j}{2}},
\end{align}
so that the matrix model $Z(D^{(2)}_{r+1},2N-1)$ \eqref{ZD} is written as
\begin{align}
&Z(D^{(2)}_{r+1},2N-1)\nonumber \\
&=\frac{1}{((2N-1)!)^2}\int
\prod_{i=1}^{2N-1}\frac{dx^{(1)}_i}{2\pi}
\prod_{i=1}^{2N-1}\frac{dx^{(r-1)}_i}{2\pi}
{\cal O}^{0,1}_{\circ\!\Leftarrow,\text{odd}}[x^{(1)}_i]
{\cal O}^{1,\cdots,r-1}_{-}[x^{(1)}_i,x^{(r-1)}_i]
{\cal O}^{r,r-1}_{\circ\!\Leftarrow,\text{odd}}[x^{(r-1)}_i].
\label{ZDoddinO}
\end{align}
This ${\cal O}_{\circ\!\Leftarrow,\text{odd}}^{0,1}[x^{(1)}_i]$ can be rewritten by following the same strategy as in section \ref{sec_buildingblocks} and using the formula \eqref{VdMlikeforsgnNodd} as
\begin{align}
&{\cal O}_{\circ\!\Leftarrow,\text{odd}}^{0,1}[x^{(1)}_i]\nonumber \\
&=\frac{1}{(2N-1)!}\int\prod_{i=1}^{2N-1}\frac{dx^{(0)}_i}{2\pi}\operatorname*{pf}_{2N\times 2N}
\begin{pmatrix}
[\langle x^{(0)}_i|\frac{-2i}{\hat{p}}|x^{(0)}_j\rangle]_{i,j}^{(2N-1)\times (2N-1)}
&
[\langle x^{(0)}_i|0\rangle\!\rangle]_{i=1}^{2N-1}\vspace{0.2cm} \\
[\langle\!\langle 0|x^{(0)}_j\rangle]_{j=1}^{2N-1}
&
0
\end{pmatrix}\nonumber \\
&\quad\times
\det\Bigl(\langle x^{(0)}_i|(\hat{\mu}^{(0)})^{\frac{1}{2}}\hat{\cal O}^{0,1}_-|x^{(0)}_j\rangle\Bigr).
\end{align}
Plugging this into \eqref{ZDoddinO}, performing the $x^{(1)}_i,x^{(r-1)}_i$-integrations using the Andréief identity \eqref{Andreief} and the $x^{(r)}_i$ integrations using the De Bruijn's formula \eqref{BruijnN+1xN+1}, we obtain
\begin{align}
&Z(D^{(2)}_{r+1};2N-1)\nonumber \\
&=\frac{1}{(2N-1)!}\int\prod_{i=1}^{2N-1}\frac{dx_i}{2\pi}\nonumber \\
&\quad\times
\operatorname*{pf}_{4N\times 4N}
{\fontsize{9.5pt}{20pt}\selectfont
\begin{pmatrix}
[
\langle x_i|
\hat{H}_{11}
|x_j\rangle]_{i,j}^{(2N-1)\times (2N-1)}&
[\langle x_i|\hat{H}_{12}|0\rangle\!\rangle]_{i=1}^{2N-1}
&[0]^{(2N-1)\times (2N-1)}
&[0]^{(2N-1)\times 1}\\
[-\langle\!\langle 0|\hat{H}_{21}|x_j\rangle]_{j=1}^{2N-1}
&0
&[0]^{1\times (2N-1)}
&0\\
[0]^{(2N-1)\times (2N-1)}
&[0]^{(2N-1)\times 1}
&[\langle x_i|\frac{-2i}{\hat{p}}|x_j\rangle]_{i,j}^{(2N-1)\times (2N-1)}
&[\langle x_i|0\rangle\!\rangle]_{i=1}^{2N-1}\\
[0]^{1\times (2N-1)}
&0
&[-\langle\!\langle 0|x_j\rangle]_{j=1}^{2N-1}
&0
\end{pmatrix}
}\nonumber \\
&=-\frac{1}{(2N-1)!}\int\prod_{i=1}^{2N-1}\frac{dx_i}{2\pi}\nonumber \\
&\quad\times
\operatorname*{pf}_{4N\times 4N}
{\fontsize{9.5pt}{20pt}\selectfont
\begin{pmatrix}
[
\langle x_i|
\hat{H}_{11}
|x_j\rangle
]_{i,j}^{(2N-1)\times (2N-1)}
&[0]^{(2N-1)\times (2N-1)}
&[\langle x_i|\hat{H}_{12}|0\rangle\!\rangle]_{i=1}^{2N-1}
&[0]^{(2N-1)\times 1}\\
[0]^{(2N-1)\times (2N-1)}
&[\langle x_i|\frac{-2i}{\hat{p}}|x_j\rangle]_{i,j}^{(2N-1)\times (2N-1)}
&[0]^{(2N-1)\times 1}
&[\langle x_i|0\rangle\!\rangle]_{i=1}^{2N-1}\\
[-\langle\!\langle 0|\hat{H}_{21}|x_j\rangle]_{j=1}^{2N-1}
&[0]^{1\times (2N-1)}
&0
&0\\
[0]^{1\times (2N-1)}
&[-\langle\!\langle 0|x_j\rangle]_{j=1}^{2N-1}
&0
&0
\end{pmatrix}
}
\label{ZD(2)oddNstep2}
\end{align}
with
\begin{subequations}
\begin{align}
&\hat{H}_{11}
=
(\hat{\mu}^{(0)})^{\frac{1}{2}}
\hat{\cal O}^{0,\cdots,r}_-
(\hat{\mu}^{(r)})^{\frac{1}{2}}
\frac{-2i}{\hat{p}}
(\hat{\mu}^{(r)})^{\frac{1}{2}}
\hat{\cal O}^{r,\cdots,0}_-
(\hat{\mu}^{(0)})^{\frac{1}{2}},\\
&\hat{H}_{12}=(\hat{\mu}^{(0)})^{\frac{1}{2}}\hat{\cal O}^{0,\cdots,r}_-(\hat{\mu}^{(r)})^{\frac{1}{2}}=
\hat{\mu}^{(0)}
\frac{1}{2\cosh\frac{\hat{p}}{2}}
\cdots
\frac{1}{2\cosh\frac{\hat{p}}{2}}
\hat{\mu}^{(r)},\\
&\hat{H}_{21}=(\hat{\mu}^{(r)})^{\frac{1}{2}}\hat{\cal O}^{r,\cdots,0}_-(\hat{\mu}^{(0)})^{\frac{1}{2}}=(\hat{H}_{12})^{\rmT}.
\end{align}
\end{subequations}
Here in the second expression in \eqref{ZD(2)oddNstep2} we have exchanged the $2N$-th row/column with $(2N+1),\cdots,(4N-1)$-th row/columns, which produces the overall minus sign.

Now let us define the generating function of $Z(D^{(2)}_{r+1};2N-1)$ as
\begin{align}
&\Xi_{\text{odd}}(D^{(2)}_{r+1};u)\nonumber \\
&=\sum_{N=1}^\infty u^{N-1}Z(D^{(2)}_{r+1};2N-1)\nonumber \\
&=-\sum_{M\in 2\mathbb{N}-1}(-u)^{\frac{M-1}{2}}\frac{(-1)^{\frac{M(M-1)}{2}}}{M!}\int\prod_{i=1}^M\frac{dx_i}{2\pi}\nonumber \\
&\quad\times \operatorname*{pf}_{(2M+2)\times (2M+2)}
\begin{pmatrix}
[
\langle x_i|
\hat{H}_{11}
|x_j\rangle
]_{i,j}^{M\times M}
&[0]^{M\times M}
&[\langle x_i|\hat{H}_{12}|0\rangle\!\rangle]_{i=1}^M
&[0]^{M\times 1}\vspace{0.2cm} \\
[0]^{M\times M}
&[\langle x_i|\frac{-2i}{\hat{p}}|x_j\rangle]_{i,j}^{M\times M}
&[0]^{M\times 1}
&[\langle x_i|0\rangle\!\rangle]_{i=1}^M\vspace{0.2cm} \\
[-\langle\!\langle 0|\hat{H}_{21}|x_j\rangle]_{j=1}^M
&[0]^{1\times M}
&0
&0\vspace{0.2cm} \\
[0]^{1\times M}
&[-\langle\!\langle 0|x_j\rangle]_{j=1}^M
&0
&0
\end{pmatrix},
\label{XiD(2)oddNstep1}
\end{align}
where in the second expression we have substituted $Z(D^{(2)}_{r+1};2N-1)$ \eqref{ZD(2)oddNstep2} and denoted $2N-1=M$.
Since the Pfaffian in \eqref{XiD(2)oddNstep1} vanishes trivially for even $M$,\footnote{
This can be seen by considering, for arbitrary non-degenerate anti-symmetric $n\times n$ matrices $A$ and $B$ and arbitrary $n$-component vectors $v$ and $w$,
\begin{align}
\left(
\operatorname{pf}
\begin{pmatrix}
A&0&v&0\\
0&B&0&w\\
-v^{\rmT}&0&0&0\\
0&-w^{\rmT}&0&0
\end{pmatrix}
\right)^2
=
\det
\begin{pmatrix}
A&0&v&0\\
0&B&0&w\\
-v^{\rmT}&0&0&0\\
0&-w^{\rmT}&0&0
\end{pmatrix}
=\det A\det B(v^{\rmT}A^{-1}v)(w^{\rmT}B^{-1}w)=0.
\end{align}
This quantity vanishes because $v^{\rmT} A^{-1}v=w^{\rmT} B^{-1}w=0$ due to the antisymmetric property of $A$ and $B$.
}
we can replace the range of the summation from $2\mathbb{N}-1$ to $\mathbb{Z}_{\ge 0}$.
After that, we can rewrite $\Xi_\text{odd}(D^{(2)}_{r+1};u)$ by using the Fredholm Pfaffian formula with rank deformation \eqref{FredholmPfaffianrankdeform} as
\begin{align}
\Xi_{\text{odd}}(D^{(2)}_{r+1};u)=-(-u)^{-\frac{1}{2}}\sqrt{\operatorname{Det}\begin{pmatrix}
(-u)^{\frac{1}{2}}\hat{H}_{11}
&1
&(-u)^{\frac{1}{2}}\hat{H}_{12}|0\rangle\!\rangle
&0\\
-1
&(-u)^{\frac{1}{2}}\frac{-2i}{\hat{p}}
&0
&(-u)^{\frac{1}{2}}|0\rangle\!\rangle\\
-\langle\!\langle 0|\hat{H}_{21}
&0
&0
&0\\
0
&-\langle\!\langle 0|
&0
&0
\end{pmatrix}
}.
\end{align}
By further performing elementary row/column transformations, we obtain
\begin{align}
\Xi_{\text{odd}}(D^{(2)}_{r+1};u)=-\sqrt{
\det\begin{pmatrix}
(-u)^{\frac{1}{2}}m_{11}
&m_{12}\\
m_{21}
&(-u)^{\frac{1}{2}}m_{22}
\end{pmatrix}
}\sqrt{\operatorname{Det}(1+u\hat{\rho}_{\text{odd}}(D^{(2)}_{r+1}))}
\label{XioddD(2)r+1final}
\end{align}
where
\begin{align}
&\hat{\rho}_{\text{odd}}(D^{(2)}_{r+1})=-\hat{H}_{11}\frac{-2i}{\hat{p}}
=
\hat{\mu}^{(0)}\frac{1}{2\cosh\frac{\hat{p}}{2}}\cdots\frac{1}{2\cosh\frac{\hat{p}}{2}}\hat{\mu}^{(r)}
\frac{2}{\hat{p}}
\hat{\mu}^{(r)}\frac{1}{2\cosh\frac{\hat{p}}{2}}\cdots\frac{1}{2\cosh\frac{\hat{p}}{2}}\hat{\mu}^{(0)}
\frac{2}{\hat{p}},
\end{align}
which coincides with $\hat{\rho}(D^{(2)}_{r+1})$ \eqref{rhohatD} up to a similarity transformation.

The matrix elements $m_{ab}$ of the $2\times 2$ matrix are
\begin{subequations}
\begin{align}
&m_{11}=\langle\!\langle 0|\hat{H}_{21}\frac{-2i}{\hat{p}}\frac{1}{1+u\hat{\rho}_{\text{odd}}(D^{(2)}_{r+1})}\hat{H}_{12}|0\rangle\!\rangle=0,\\
&m_{12}=-\langle\!\langle 0|\hat{H}_{21}\frac{1}{1+u\hat{\rho}_{\text{odd}}(D^{(2)}_{r+1})^{\rmT}}|0\rangle\!\rangle=-m_{21},\\
&m_{21}=\langle\!\langle 0|\frac{1}{1+u\hat{\rho}_{\text{odd}}(D^{(2)}_{r+1})}\hat{H}_{12}|0\rangle\!\rangle,\\
&m_{22}=\langle\!\langle 0|\frac{1}{1+u\hat{\rho}_{\text{odd}}(D^{(2)}_{r+1})}\hat{H}_{11}|0\rangle\!\rangle=0.
\end{align}
\end{subequations}
Here $m_{11}$ and $m_{22}$ vanish since their expansion coefficients take the form of $\langle\!\langle 0|\hat{\cal O}|0\rangle\!\rangle$ with $\hat{\cal O}^{\rmT}=-\hat{\cal O}$.
The relation between $m_{12}$ and $m_{21}$ can also be shown by taking the transpose of $m_{12}$.
Plugging these into \eqref{XiD(2)oddNstep1}, we finally obtain
\begin{align}
\Xi_{\text{odd}}(D^{(2)}_{r+1};u)=
\langle\!\langle 0|\frac{1}{1+u\hat{\rho}_{\text{odd}}(D^{(2)}_{r+1})}
\hat{\mu}^{(0)}
\frac{1}{2\cosh\frac{\hat{p}}{2}}
\cdots
\frac{1}{2\cosh\frac{\hat{p}}{2}}
\hat{\mu}^{(r)}
|0\rangle\!\rangle
\sqrt{\operatorname{Det}(1+u\hat{\rho}_{\text{odd}}(D^{(2)}_{r+1}))}.
\end{align}
Note that the overall sign can be confirmed by comparing the coefficients of $u^0$ of both sides, which is $Z(D^{(2)}_{r+1},1)=\langle\!\langle
0|
\hat{\mu}^{(0)}
\frac{1}{2\cosh\frac{\hat{p}}{2}}
\cdots
\frac{1}{2\cosh\frac{\hat{p}}{2}}
\hat{\mu}^{(r)}
 |0\rangle\!\rangle$.

\bibliographystyle{utphys}
\bibliography{ref_draft.bib}

\end{document}